\documentclass[12pt]{article}

\usepackage{graphics,graphpap, myfeynarts}
\usepackage{epsfig,graphicx}
\usepackage{fancybox,amsmath}

\newcommand{\bea}{\begin{eqnarray}}
\newcommand{\eea}{\end{eqnarray}}
\newcommand{\beq}{\begin{equation}}
\newcommand{\eeq}{\end{equation}}

\def\msbar{\ifmmode{\overline{\rm MS}} \else{$\overline{\rm MS}$} \fi}
\def\drbar{\ifmmode{\overline{\rm DR}} \else{$\overline{\rm DR}$} \fi}
\def\sf{\ifmmode{\tilde{f}} \else{$\tilde{f}$} \fi}
\def\st{\ifmmode{\tilde{t}} \else{$\tilde{t}$} \fi}
\def\sb{\ifmmode{\tilde{b}} \else{$\tilde{b}$} \fi}
\def\sq{\ifmmode{\tilde{q}} \else{$\tilde{q}$} \fi}
\def\sg{\ifmmode{\tilde{g}} \else{$\tilde{g}$} \fi}
\def\bbar{\ifmmode{\bar{b}} \else{$\bar{b}$} \fi}
\def\tbar{\ifmmode{\bar{t}} \else{$\bar{t}$} \fi}
\def\qbar{\ifmmode{\bar{q}} \else{$\bar{q}$} \fi}
\def\ksla{{k \hspace{-2mm} \slash}}

\def\sq              {{\tilde q}}

\def\sf              {{\tilde f}}

\newcommand\bvec{\left( \begin{array}{c}}
\newcommand\evec{\end{array}\right)}

\newcommand\bmat{\left( \begin{array}{cc}}
\newcommand\emat{\end{array}\right)}

\renewcommand\d{\delta}

\def\ksla{{k \hspace{-2.2mm} \slash}}

\newcommand\cchp{{\tilde\chi^+}}
\newcommand\cchm{{\tilde\chi^-}}
\newcommand\cchpm{{\tilde\chi^\pm}}

\newcommand\cch{{\tilde\chi^0}}
\newcommand\nch{{\tilde\chi}}

\def\chp             {\tilde \chi^+}

\def\onehf           {{\textstyle \frac{1}{2}}}

\renewcommand\d{\delta}
\newcommand\D{\Delta}
\renewcommand\a{\alpha}
\renewcommand\b{\beta}
\newcommand\g{\gamma}
\newcommand\G{\Gamma}
\newcommand\e{\epsilon}
\newcommand\la{\lambda}
\renewcommand\t{\theta}
\newcommand\s{\sigma}

\newcommand\ds{{\rm d}\sigma}

\newcommand\ReTilde{\widetilde{\rm Re}}

\renewcommand\Re{{\rm Re}}
\renewcommand\Im{{\rm Im}}

\newcommand{\ZN}{N}
\newcommand{\Mse}{M_{\tilde E_1}}
\newcommand{\Msl}{M_{\tilde L_1}}

\def\su{\ifmmode{\tilde{u}} \else{$\tilde{u}$} \fi}
\def\sd{\ifmmode{\tilde{d}} \else{$\tilde{d}$} \fi}

\def\non             {\nonumber}

\begin{document}

\pagestyle{empty} \vspace*{-1cm}
\begin{flushright}
  HEPHY-PUB 806/05 \\
  hep-ph/0504109
\end{flushright}

\vspace*{1.4cm}

\begin{center}
\begin{Large} \bf
Precise predictions for chargino and neutralino pair production in {\boldmath $e^+e^-$} annihilation
\end{Large}

\vspace{10mm}

{\large W.~\"Oller, H.~Eberl, W.~Majerotto}

\vspace{6mm}
\begin{tabular}{l}
 {\it Institut f\"ur Hochenergiephysik der \"Osterreichischen
 Akademie der Wissenschaften,}\\
{\it A--1050 Vienna, Austria}
\end{tabular}

\vspace{20mm}

\begin{abstract}
We present a high precision calculation of chargino and neutralino pair production at $e^+e^-$ colliders.
Within the Minimal Supersymmetric Standard Model, the full one-loop and higher order QED corrections are included. 
Special care has been taken in the definition of the Lagrangian input parameters. 
Furthermore, the proper inclusion of QED corrections and the separation of weak and QED corrections
are discussed. We show numerical results for total cross sections, as well as forward-backward and
left-right asymmetries for the SPS1a' scenario as proposed in the SPA project. The complete corrections are about 10\%
and in some cases even larger, in particular for $\cch_i\cch_j$ production with sizeable higgsino components. These corrections have 
to be taken into account in a high precision analysis.
\end{abstract}
\end{center}

\vfill

\newpage
\pagestyle{plain} \setcounter{page}{2}

\section{Introduction}
Supersymmetry predicts the existence of fermionic partners to the gauge and Higgs bosons. In the Supersymmetric Standard Model (MSSM),
one has two charginos $\cchpm_1$ and $\cchpm_2$, which are the fermionic mass eigenstates of the supersymmetric partners of the $W^\pm$
and the charged Higgs states $H^\pm_{1,2}$. There are four neutralinos $\cch_1 - \cch_4$ being the fermionic partners of the photon, $Z^0$ boson,
and the neutral Higgs boson states $H^0_{1,2}$. At tree-level, the chargino and neutralino systems depends only on the parameters $M$, $M'$, $\mu$, and
$\tan\b=v_2/v_1$, with $v_{1,2}$ the vacuum expectation values of the two neutral Higgs doublet fields.\\
It is expected that at least some of these particles will be detected at LHC or Tevatron, most likely in the cascade decays of gluinos
and squarks. The properties of charginos and neutralinos can be studied with high accuracy at a high energy linear $e^+e^-$ collider \cite{LCstudies}.
At lowest order, the above mentioned parameters can be extracted from the masses and production cross sections in $e^+e^-$ collisions
with polarized beams \cite{chaparam, neuparam}. The experimental accuracy of measurements to be obtained at a linear collider will be so high
that it is necessary to incorporate effects beyond leading order in the theoretical calculations in order to match the experimental precision.
Beyond tree-level, the definition of the parameters involved are no more unique and depends on the renormalization scheme. Therefore, a well
defined theoretical framework has been recently proposed within the so-called SPA (SUSY Parameter Analysis) project \cite{SPA}. The proposed
``SPA convention" provides a clear base for calculating masses, mixings, decay widths and production processes. The final goal is to extract
the fundamental supersymmetry parameters from data.\\
In this paper, we treat in detail the calculation of chargino and neutralino pair production at full one-loop level within the MSSM.
Particular attention is paid to a proper inclusion of QED corrections and to a suitable separation of weak and QED corrections.
For the initial state radiation of photons, we use the structure function formalism \cite{sff}, where also non-leading log terms are included
and the soft-photon contributions are summed up to all orders in perturbation theory. We adopt the SPA convention, in which the SUSY input
parameters are defined in the $\drbar$ scheme at the scale $Q$=1 TeV. The parameters are then translated to the on-shell scheme in which
the calculations are performed. The result for the observables (cross sections, asymmetries, etc.) is therefore independent of the
renormalization scheme up to higher order corrections. The masses of the SUSY particles and the Higgs bosons are of course defined as
physical pole masses.\\
This paper is organized in the following way: Section~\ref{Notations} introduces the used notation and conventions. 
In section~\ref{Tree-level} the tree-level cross section is given. Section~\ref{One-loop corrections} contains the calculation of virtual
corrections and describes in detail the used renormalization scheme. In section~\ref{QED corrections} the inclusion of the QED corrections
and the separation of the weak and QED part is discussed. Section~\ref{Numerical results} shows the numerical analysis for the SPS1a' scenario,
proposed within the SPA project. Finally, section~\ref{Conclusions} summarizes this paper.  

\section{Notations}\label{Notations}
In supersymmetric extensions of the SM, the fermionic superpartners of the vector and Higgs bosons, the gauginos and higgsinos,
can mix and form the mass eigenstates of the neutralinos and the charginos. In the MSSM the mixing
is defined by the two mass matrices, which are non-diagonal after the gauge symmetry breaking. 
The symmetric tree-level mass matrix of the neutral $\psi_i^0=(-i\la',-i\la^3,\psi^1_{H_1},\psi^2_{H_2})$ Weyl states
\begin{equation}
    Y = \left(\begin{array}{cccc}
M' & 0 & - m_Z \sin\theta_W \cos\beta & m_Z \sin\theta_W
\sin\beta\\ 0 & M & m_Z \cos\theta_W \cos\beta & -m_Z \cos\theta_W
\sin\beta\\ - m_Z \sin\theta_W \cos\beta & m_Z \cos\theta_W
\cos\beta & 0 & -\mu\\ m_Z \sin\theta_W \sin\beta & -m_Z
\cos\theta_W \sin\beta & -\mu & 0
\end{array}\right)
 \label{neumat}
\end{equation}
can be diagonalized by the unitary matrix $\ZN$
\begin{equation}
\mbox{diag}(m_{\tilde\chi_1^0},\,m_{\tilde\chi_2^0},\,m_{\tilde\chi_3^0},\,m_{\tilde\chi_4^0})= \ZN^*\,Y\,\ZN^\dag 
\,.
\end{equation}
The mass eigenstates in the 4-component Majorana spinor notation $\cch_i$ are defined by
\begin{equation}
\chi_i^0=N_{ij}\psi^0_j\,,\qquad\cch_i = \left(\begin{array}{c} \chi_i^0 \\ \overline{\chi}_i^0 \end{array}\right)\,.
\end{equation}
 
The diagonalization of the chargino tree-level mass matrix
\begin{equation}
    X = \left(\begin{array}{cc}
M & \sqrt{2}\,m_W\cos\b \\ 
\sqrt{2}\,m_W\sin\b & \mu
\end{array}\right)
 \label{chmat}
\end{equation}
can be performed by two unitary matrices
\begin{equation}
\mbox{diag}(m_{\tilde\chi_1^\pm},\,m_{\tilde\chi_2^\pm})= U^*\,X\,V^\dag 
\,.
\end{equation}
Thus, the chargino mass eigenstates in Dirac form $\tilde\chi_i^\pm$ is obtained from the Weyl states 
$\psi^+=(-i\la^+,\psi^1_{H_2})$, $\psi^-=(-i\la^-,\psi^2_{H_1})$ 
by the relations
\begin{equation}
\chi_i^+=V_{ij}\psi^+_j\,,\qquad\chi_i^-=U_{ij}\psi^-_j\,,\qquad
\tilde\chi^\pm_i = \left(\begin{array}{c} \chi_i^\pm \\ \overline{\chi}_i^\mp \end{array}\right)\,.
\end{equation}
 
The spectrum of the neutralinos and charginos is specified by the soft SU(2) and U(1) gaugino mass parameters $M$ and $M'$, 
the Higgs/higgsino parameter $\mu$ and $\tan\beta$.\\
We assume no flavour mixing in the sfermion sector and further neglect the selectron L-R mixing, which is suppressed
by a factor $m_e$ in the off-diagonal selectron mass matrix elements. 
Thus, we get for the selectron masses and the sneutrino mass in terms of electroweak parameters and SUSY breaking masses at tree-level
\begin{eqnarray}
m^2_{\tilde{e}_L} &=& \Msl^2 - m_Z^2 \cos2\b \left( \onehf - \sin^2\theta_W \right) 
\\
m^2_{\tilde{e}_R} &=& \Mse^2 - m_Z^2 \cos2\b \, \sin^2\theta_W 
\\
m^2_{\tilde{\nu}_e} &=& \Msl^2 + \onehf m_Z^2 \cos2\b
\end{eqnarray}

\section{Tree-level}\label{Tree-level}
\begin{figure}[h!]
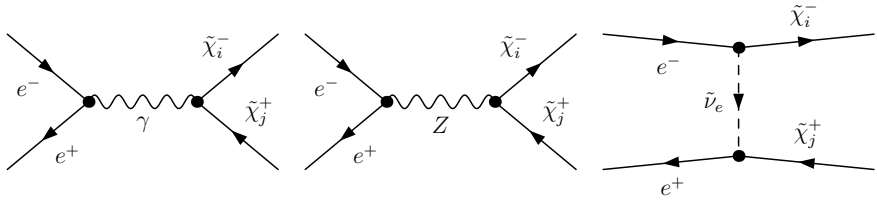

\begin{center}
\mbox{\resizebox{200mm}{!}{
\begin{feynartspicture}(750,150)(4,1)
\FADiagram{}
\FAProp(0.,15.)(6.,10.)(0.,){/Straight}{1}
\FALabel(2.48771,11.7893)[tr]{$e^-$}
\FAProp(0.,5.)(6.,10.)(0.,){/Straight}{-1}
\FALabel(3.51229,6.78926)[tl]{$e^+$}
\FAProp(20.,15.)(14.,10.)(0.,){/Straight}{-1}
\FALabel(16.4877,13.2107)[br]{$\tilde \chi_i^-$}
\FAProp(20.,5.)(14.,10.)(0.,){/Straight}{1}
\FALabel(17.5123,8.21074)[bl]{$\tilde \chi_j^+$}
\FAProp(6.,10.)(14.,10.)(0.,){/Sine}{0}
\FALabel(10.,8.93)[t]{$\gamma$}
\FAVert(6.,10.){0}
\FAVert(14.,10.){0}

\FADiagram{}
\FAProp(0.,15.)(6.,10.)(0.,){/Straight}{1}
\FALabel(2.48771,11.7893)[tr]{$e^-$}
\FAProp(0.,5.)(6.,10.)(0.,){/Straight}{-1}
\FALabel(3.51229,6.78926)[tl]{$e^+$}
\FAProp(20.,15.)(14.,10.)(0.,){/Straight}{-1}
\FALabel(16.4877,13.2107)[br]{$\tilde \chi_i^-$}
\FAProp(20.,5.)(14.,10.)(0.,){/Straight}{1}
\FALabel(17.5123,8.21074)[bl]{$\tilde \chi_j^+$}
\FAProp(6.,10.)(14.,10.)(0.,){/Sine}{0}
\FALabel(10.,8.93)[t]{$Z$}
\FAVert(6.,10.){0}
\FAVert(14.,10.){0}

\FADiagram{}
\FAProp(0.,15.)(10.,14.)(0.,){/Straight}{1}
\FALabel(4.84577,13.4377)[t]{$e^-$}
\FAProp(0.,5.)(10.,6.)(0.,){/Straight}{-1}
\FALabel(5.15423,4.43769)[t]{$e^+$}
\FAProp(20.,15.)(10.,14.)(0.,){/Straight}{-1}
\FALabel(14.8458,15.5623)[b]{$\tilde \chi_i^-$}
\FAProp(20.,5.)(10.,6.)(0.,){/Straight}{1}
\FALabel(15.1542,6.56231)[b]{$\tilde \chi_j^+$}
\FAProp(10.,14.)(10.,6.)(0.,){/ScalarDash}{1}
\FALabel(8.93,10.)[r]{$\tilde \nu_e$}
\FAVert(10.,14.){0}
\FAVert(10.,6.){0}

\end{feynartspicture}
}}
\caption[chatree]
{Tree-level chargino production}
 \label{fig:chatree}
\end{center}
\end{figure}

\begin{figure}[h!]
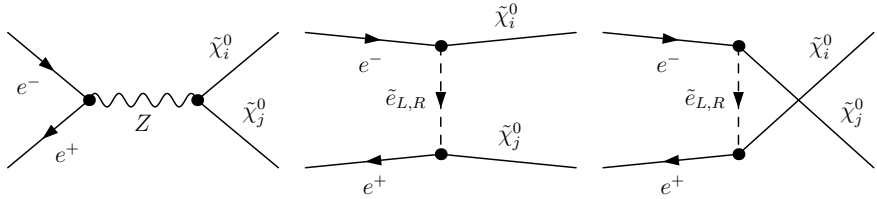

\begin{center}
\mbox{\resizebox{200mm}{!}{
\begin{feynartspicture}(750,150)(4,1)
\FADiagram{}
\FAProp(0.,15.)(6.,10.)(0.,){/Straight}{1}
\FALabel(2.48771,11.7893)[tr]{$e^-$}
\FAProp(0.,5.)(6.,10.)(0.,){/Straight}{-1}
\FALabel(3.51229,6.78926)[tl]{$e^+$}
\FAProp(20.,15.)(14.,10.)(0.,){/Straight}{0}
\FALabel(16.6478,13.0187)[br]{$\tilde \chi_i^0$}
\FAProp(20.,5.)(14.,10.)(0.,){/Straight}{0}
\FALabel(17.3522,8.01869)[bl]{$\tilde \chi_j^0$}
\FAProp(6.,10.)(14.,10.)(0.,){/Sine}{0}
\FALabel(10.,8.93)[t]{$Z$}
\FAVert(6.,10.){0}
\FAVert(14.,10.){0}

\FADiagram{}
\FAProp(0.,15.)(10.,14.)(0.,){/Straight}{1}
\FALabel(4.84577,13.4377)[t]{$e^-$}
\FAProp(0.,5.)(10.,6.)(0.,){/Straight}{-1}
\FALabel(5.15423,4.43769)[t]{$e^+$}
\FAProp(20.,15.)(10.,14.)(0.,){/Straight}{0}
\FALabel(14.8706,15.3135)[b]{$\tilde \chi_i^0$}
\FAProp(20.,5.)(10.,6.)(0.,){/Straight}{0}
\FALabel(15.1294,6.31355)[b]{$\tilde \chi_j^0$}
\FAProp(10.,14.)(10.,6.)(0.,){/ScalarDash}{1}
\FALabel(8.93,10.)[r]{$\tilde e_{L,R}$}
\FAVert(10.,14.){0}
\FAVert(10.,6.){0}

\FADiagram{}
\FAProp(0.,15.)(10.,14.)(0.,){/Straight}{1}
\FALabel(4.84577,13.4377)[t]{$e^-$}
\FAProp(0.,5.)(10.,6.)(0.,){/Straight}{-1}
\FALabel(5.15423,4.43769)[t]{$e^+$}
\FAProp(20.,15.)(10.,6.)(0.,){/Straight}{0}
\FALabel(16.98,13.02)[br]{$\tilde \chi_i^0$}
\FAProp(20.,5.)(10.,14.)(0.,){/Straight}{0}
\FALabel(17.52,8.02)[bl]{$\tilde \chi_j^0$}
\FAProp(10.,14.)(10.,6.)(0.,){/ScalarDash}{1}
\FALabel(9.03,10.)[r]{$\tilde e_{L,R}$}
\FAVert(10.,14.){0}
\FAVert(10.,6.){0}

\end{feynartspicture}
}}
\caption[neutree]
{Tree-level neutralino production}
 \label{fig:neutree}
\end{center}
\end{figure}

The tree-level pair production processes for charginos, Fig.~\ref{fig:chatree}, and neutralinos, Fig.~\ref{fig:neutree}, 
\begin{eqnarray}
&&e^-(p_1)\,e^+(p_2) \rightarrow \cchm_i(k_1)\,\cchp_j(k_2)  \qquad (i,j = 1,2)\,, \nonumber \\  
&&e^-(p_1)\,e^+(p_2) \rightarrow \cch_i\hspace{0.8mm}(k_1)\,\cch_j\hspace{0.7mm}(k_2)\qquad  (i,j = 1,2,3,4)\,, \nonumber
\end{eqnarray}
have been already extensively discussed in the literature \cite{neuparam, chaparam}.
After a Fierz transformation of the sfermion exchange channels, where the electron mass is neglected in the Yukawa couplings, 
the tree-level matrix-elements can be written in terms of four independent helicity amplitudes
\begin{eqnarray}
{\cal M}^{0,\pm}_{\a\b} &=& i\frac{e^2}{s} Q^{0,\pm}_{\a\b}\left[ \bar{v}(p_2)\,\g_\mu P_\a\,u(p_1)\right] \left[\bar{u}(k_1)\,\g^\mu P_\b\,v(k_2)\right]
\,,\hspace{0.7cm}\{\a,\b\}\,\,\epsilon\,\,\{L,R\}\,.
\end{eqnarray}
The bilinear charges $Q^\pm_{\a\b}$ for charginos and $Q^0_{\a\b}$ for neutralinos are 
\begin{center}
\begin{tabular}{ll}
$Q^\pm_{LL} = \d_{ij}+D_Z\,C_L\,{\cal U}_{ij}\,,\qquad $ & 
$Q^\pm_{LR} = \d_{ij}+D_Z\,C_L\,{\cal V}_{ij}+D_{\tilde\nu}\,\widetilde{{\cal V}}_{ij}\,,$\\
$Q^\pm_{RL} = \d_{ij}+D_Z\,C_R\,{\cal U}_{ij}\,,\qquad $ &
$Q^\pm_{RR} = \d_{ij}+D_Z\,C_R\,{\cal V}_{ij}\,,$\\ \\
$Q^0_{LL} = D_Z\,C_L\,{\cal N}_{ij}-D_{u,L}\,{\cal L}_{ij}\,,\qquad $ & 
$Q^0_{LR} = -D_Z\,C_L\,{\cal N}^*_{ij}+D_{t,L}\,{\cal L}^*_{ij}\,,$\\
$Q^0_{RL} = D_Z\,C_R\,{\cal N}_{ij}+D_{t,R}\,{\cal R}_{ij}\,,\qquad $ &
$Q^0_{RR} = -D_Z\,C_R\,{\cal N}^*_{ij}-D_{u,R}\,{\cal R}^*_{ij}\,.$
\end{tabular}
\end{center}
The first index
denotes the chirality of the $e^\pm$ current, the second one of the $\nch_{i,j}$ current.  
We have introduced the projection operators $P_{L/R}=\onehf (1\mp\g_5)$, the kinematical variables
\begin{equation}
s=(p_1+p_2)^2\,,\quad t=(p_1-k_1)^2 \quad {\rm and} \quad u=(p_1-k_2)^2\,,
\end{equation}
the normalized propagators
\begin{equation}
D_Z=\frac{s}{s-m_Z^2}\,,\quad D_{\tilde\nu} = \frac{s}{t-m_{\tilde{\nu}^2}}\,,\quad  D_{t,L/R} = \frac{s}{t-m^2_{\tilde{e}_{L/R}}}\,,\quad
D_{u,L/R} = \frac{s}{u-m^2_{\tilde{e}_{L/R}}}\,,
\end{equation}
and the coupling matrices
\begin{eqnarray}
&C_L=(s_W^2-\onehf)/(s_W^2\,c_W^2)\,,\quad C_R=1/c_W^2\,,&\nonumber\\
&{\cal U}_{ij} = (s_W^2\,\d_{ij} -U_{i1}U_{j1}^*-\onehf\,U_{i2}U_{j2}^*)\,,\quad {\cal V}_{ij} = {\cal U}_{ij}(U \rightarrow V)\,,&\nonumber\\
&\widetilde{{\cal V}}_{ij}=V_{i1}^*V_{j1}/(2\,s_W^2)\,,\quad {\cal N}_{ij} = (N_{i3}N_{j3}^*-N_{i4}N_{j4}^*)/2\,,&\nonumber\\
&{\cal L}_{ij} = (N_{i2}c_W+N_{i1}s_W)(N_{j2}^*c_W+N_{j1}^*s_W)/(4\,s_W^2c_W^2)\,,\quad
 {\cal R}_{ij} = N_{i1}N_{j1}^*/c_W^2\,.&
\end{eqnarray}
Summing up the final state helicities ${\cal M}^{0,\pm}_{\a}=\sum_{\b=\pm}{\cal M}^{0,\pm}_{\a\b}$, the integrated tree-level cross section for
polarized beams reads (for $m_e\rightarrow 0$)
\begin{eqnarray}
\int\ds^{\rm tree} &=& \frac{\la^{1/2}(s,m_{\nch_i},m_{\nch_j})}{64\pi^2s^2}
\int {\rm d}\Omega\sum_{\a=\pm}\frac{1}{4}(1+\a\,\xi_-)(1-\a\,\xi_+)|{\cal M}^{0,\pm}_{\a}|^2\,,
\end{eqnarray}
with $\xi_{\pm}$ the degrees of polarization of the $e^\pm$ beams, and $\la(x,y,z):=(x-y-z)^2-4yz$.
 
\section{One-loop corrections}\label{One-loop corrections}
\begin{figure}[h!]
\begin{center}
\hspace{-1cm}\mbox{\resizebox{150mm}{!}{\includegraphics{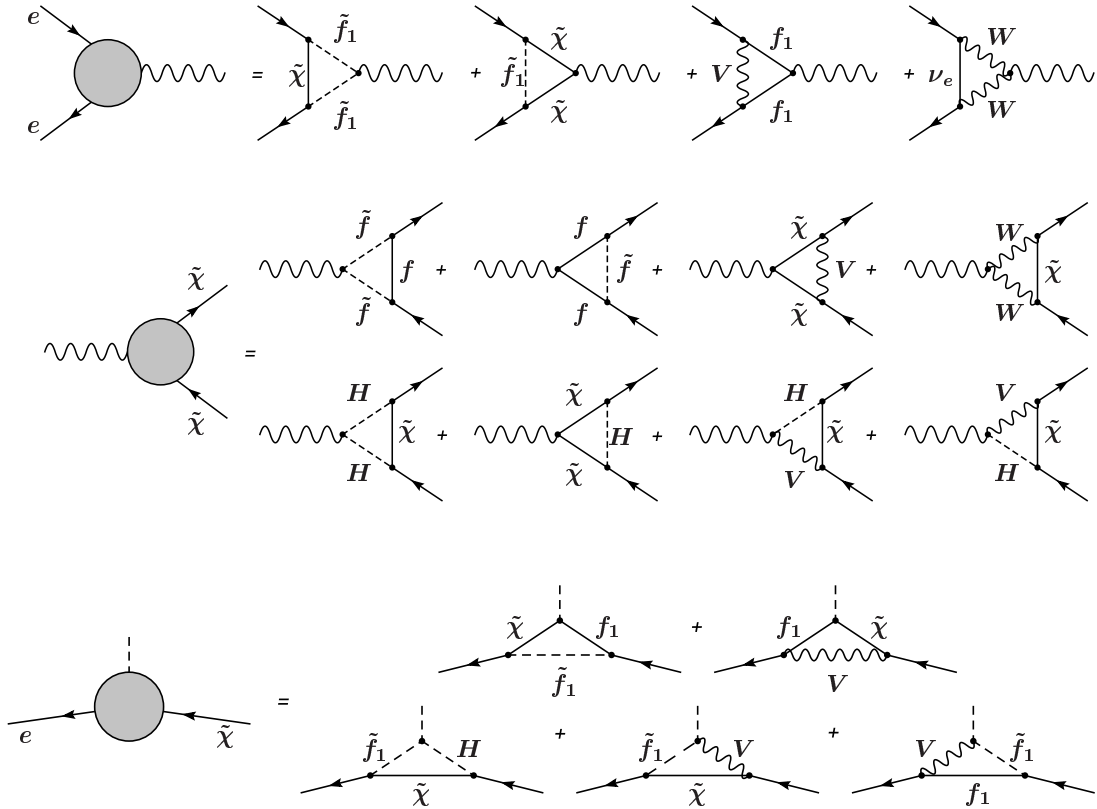}}}
\end{center}
\caption{Generic diagrams for the different vertex corrections. Seven different classes of particles are introduced:
$f$ for all SM fermions, $f_1$ for $e$ and $\nu_e$ only, $\tilde{f}$ for all sfermions, $\tilde{f}_1$ for $\tilde{e}_{L,R}$ and $\tilde{\nu}_e$ only,
$\nch$ for neutralinos and charginos, $V$ for vector bosons, and $H$ for Higgs and Goldstone bosons.   
 }
\label{fig:vertexcorr}
\end{figure}

\begin{figure}[h!]
\begin{center}
\hspace{-1cm}\mbox{\resizebox{150mm}{!}{\includegraphics{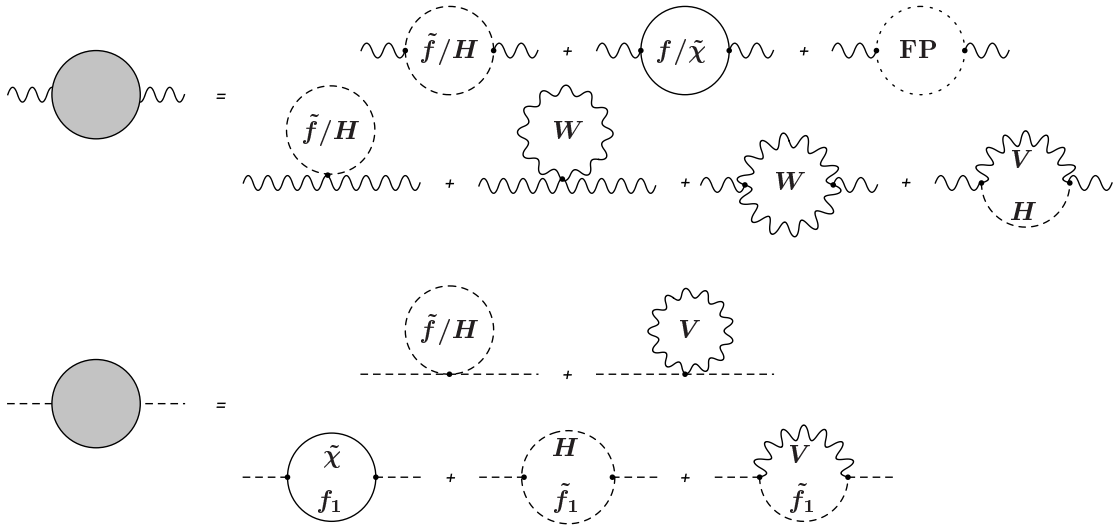}}}
\end{center}
\caption{Generic self-energy diagrams for the s-channel vector and the t/u channel slepton propagators.
FP denotes the class of Faddeev-Popov ghosts. All other particle classes are the same as in Fig.~\ref{fig:vertexcorr}.}
\label{fig:selfcorr}
\end{figure}

\begin{figure}[h!]
\begin{center}
\hspace{-1cm}\mbox{\resizebox{150mm}{!}{\includegraphics{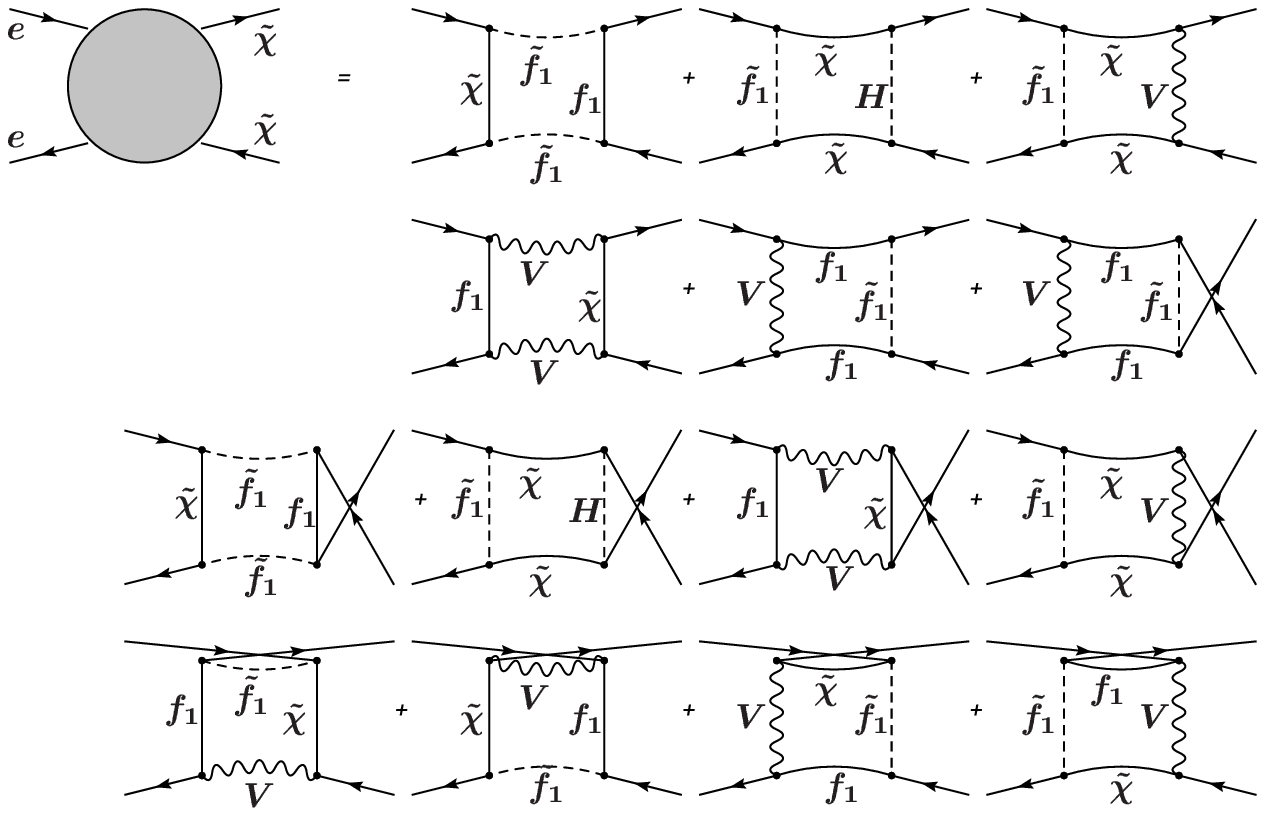}}}
\end{center}
\caption{Generic box diagrams. The notation is taken over from Fig.~\ref{fig:vertexcorr}.}
\label{fig:boxcorr}
\end{figure}

For a high precision analysis of the neutralino and chargino sector, the inclusion of higher order corrections is mandatory.
An appropriate regularization scheme to preserve supersymmetry (at least at one-loop level) is dimensional reduction ($\drbar$).
In the following elaborate calculations, a large number of Feynman diagrams are involved. This makes it necessary
to use an appropriate computer algebra tool. The FeynArts 3.2 and FormCalc 4 \cite{feyn} packages are
adopted for the generation of diagrams and amplitudes. This is performed in the $\xi=1$ Feynman-t'Hooft gauge. 
We have further integrated our renormalization scheme discussed in the next section into these programs.
The computation of the one-loop integrals is based on the packages LoopTools and FF \cite{loopFF}.\\
The virtual corrections for polarized beams can be written in the form
\begin{eqnarray}
\int\ds^{\rm virt} = \frac{\la^{1/2}(s,m_{\nch_i},m_{\nch_j})}{64\pi^2s^2}
\int {\rm d}\Omega\sum_{\a=\pm}\frac{1}{4}(1+\a\,\xi_-)(1-\a\,\xi_+)
\,2\,\Re\,\{({\cal M}_0^\a)^\dag{\cal M}_1^\a\}\,.
\end{eqnarray}
The one-loop matrix element ${\cal M}_1^\a$ consists of all possible vertex corrections (Fig.~\ref{fig:vertexcorr}), 
self-energy (Fig.~\ref{fig:selfcorr}) and box (Fig.~\ref{fig:boxcorr}) diagrams and the corresponding counter terms.
We again neglect the electron mass $m_e$ wherever possible.
\begin{figure}[h!]
\begin{center}
\hspace{-2cm}\mbox{\resizebox{120mm}{!}{\includegraphics{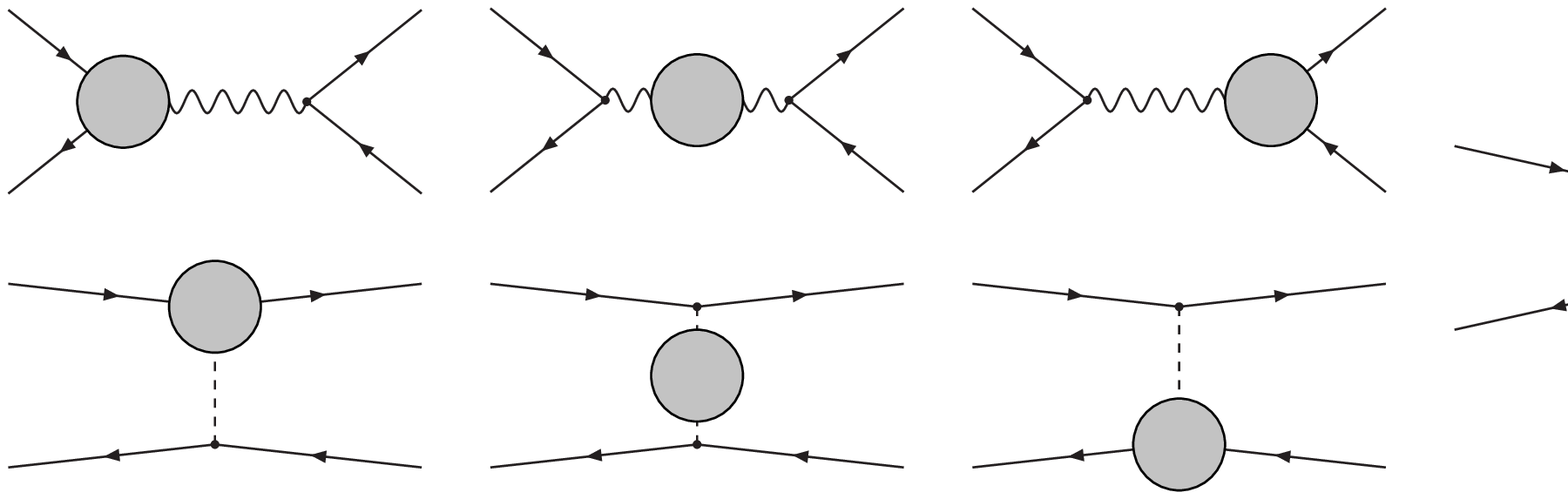}}}\,\,
\end{center}
\caption{Generic virtual diagrams. The virtual corrections are structured into vertex, propagator and box contributions.}
\label{fig:gencorr}
\end{figure}
\subsection{Renormalization scheme}
\subsubsection{Gauge sector}\label{Gauge sector}
Since the minimal supersymmetric extension of the SM does not introduce further gauge couplings, the renormalization used 
for the SM parameters can be taken from \cite{Denner}. The quark mixing matrix is assumed to be diagonal. 
The bare parameters are split into renormalized parameters and their counter terms.
\begin{eqnarray}
m_W^2\rightarrow m_W^2+\d m_W^2\,,&&\quad m_Z^2\rightarrow m_Z^2+\d m_Z^2\,,\\
f_L\rightarrow (1+\onehf\d Z_L)f_L\,,&&\quad f_R\rightarrow (1+\onehf\d Z_R)f_R\,,
\end{eqnarray}
\begin{eqnarray}
\left(\!
\begin{array}{c}
 Z \\ A
     \end{array}\!\right) & \rightarrow & \left(\begin{array}{cc}
1+\onehf\d Z_{ZZ} & \onehf\d Z_{ZA} \\ \onehf\d Z_{AZ} & 1+\onehf\d Z_{AA}
     \end{array}\right)\left(\!
\begin{array}{c}
 Z \\ A
     \end{array}\!\right)\,.
\end{eqnarray}
Since we assume no flavour mixing, no off-diagonal wave function counter terms for the SM fermions have to be introduced. 
The renormalization conditions that the on-shell masses are the real parts of the poles of the corresponding propagators and that the renormalized
fields are properly normalized, yield the following results for the counter terms.
\begin{eqnarray}\label{Vct}
\d m_Z^2~&=&~ \ReTilde\,\Pi_T^{ZZ}(m_Z^2)\,,\;\qquad \d m_W^2 = \ReTilde\,\Pi_T^{WW}(m_W^2)\,,\\
\d Z_{VV}~&=&~ - \ReTilde\,\dot\Pi_T^{VV}(m_V^2)\,,\;\quad V=A,\,Z,\,W, \\
\d Z_{AZ}~&=&~ - \frac{2\,\ReTilde\,\Pi_T^{AZ}(m_Z^2)}{m_Z^2}\,,\;\quad \d Z_{ZA}~=~\frac{2\,\ReTilde\,\Pi_T^{AZ}(0)}{m_Z^2}\,,
\end{eqnarray}
with the transverse self-energies $\Pi_T$ and $\dot\Pi (m^2)=\left[\frac{\partial}{\partial k^2}\Pi(k^2)\right]_{k^2=m^2 }$. 
Please note that $\ReTilde$ only takes the real part of the loop integrals and does not affect the possibly complex couplings.
The weak mixing angle is fixed by the usual on-shell condition \mbox{$s_W^2=1-m_W^2/m_Z^2$}.
For the SM fermions we find the wave function counter terms
\begin{eqnarray}\label{Fct}
\d Z^L ~&=&~ \ReTilde \bigg[
-\Pi^{L} (m_f^2)- m_f^2(\dot\Pi^{L} (m_f^2)+ \dot\Pi^{R}
(m_f^2))+\frac{1}{2 m_f}(\Pi^{SL} (m_f^2)-\Pi^{SR}
(m_f^2))\nonumber\\ && 
\hspace{2cm} - m_f(\dot\Pi^{SL}(m_f^2)+\dot\Pi^{SR} (m_f^2))\bigg],\\
\d Z^R~&=&~\d Z^L(L \leftrightarrow R)\,,
\end{eqnarray}
where we used the decomposition 
\begin{eqnarray}
\Pi(k^2)&=&\ksla P_L\Pi^L(k^2)+\ksla P_R\Pi^R(k^2)+P_L\Pi^{SL}(k^2)+P_R\Pi^{SR}(k^2)\,.
\end{eqnarray}
For the electric charge defined in the Thomson limit $\a\equiv e^2/(4\pi)\simeq 1/137.036$, one obtains
\begin{eqnarray}\label{deltaeSM}
\frac{\delta e}{e} = -\frac{1}{2}\delta Z_{AA} -\frac{s_W}{2c_W}\delta Z_{ZA}\,.
\end{eqnarray}
However, this definition leads to large radiative corrections for processes at the GeV or TeV scale.
Furthermore, the hadronic contributions are substantially modified by low energy strong interaction effects,
making the perturbatively obtained result unreliable.
Therefore, we use two different improved schemes.
In the first scheme, we start from the effective
$\overline{\rm MS}$ value at the $Z$ pole, $\a \equiv
\a^{\rm eff}(m_{\scriptscriptstyle Z})|_{\overline{\rm MS}} \simeq 1/127.7$, where 
only the leptonic and light quark contributions are included.
This results in the counter term \cite{deleMZ}
\begin{eqnarray}\non
  \frac{\d e}{e} &=& \frac{1}{(4\pi)^2}\,\frac{e^2}{6} \Bigg[
  \,4 \sum_f N_C^f\, e_f^2 \bigg(\D + \log\frac{Q^2}{x_f^2} \bigg)
  + \sum_{\sf} \sum_{m=1}^2 N_C^f\, e_f^2
  \bigg( \D + \log\frac{Q^2}{m_{\sf_m}^2} \bigg)
  \\ \non
  && \hspace{18mm}
  +4 \sum_{k=1}^2 \bigg( \D + \log\frac{Q^2}{m_{\chp_k}^2} \bigg)
  + \bigg( \D + \log\frac{Q^2}{m_{H^+}^2} \bigg)
  - 21 \bigg( \D + \log\frac{Q^2}{m_{\scriptscriptstyle W}^2} \bigg)
  \Bigg] \,,
  \\
\end{eqnarray}
with $x_f = m_{\scriptscriptstyle Z} \ \forall\ m_f <
m_{\scriptscriptstyle Z}$ and $x_t = m_t$.  $N_C^f$ is the colour
factor, $N_C^f = 1, 3$ for (s)leptons and (s)quarks, respectively.
$\D$ denotes the UV divergence factor, \mbox{$\D = 2/\epsilon - \g_{\rm E}
+ \log 4\pi$}.\\
The second procedure we call the $G_F$ scheme, with $\a\equiv\sqrt{2}s_W^2m_W^2G_F/\pi \simeq 1/132.5$ and
\begin{eqnarray}\label{deltaedeltar}
\frac{\delta e}{e} = -\frac{1}{2}\delta Z_{AA} -\frac{s_W}{2c_W}\delta Z_{ZA} - \frac{1}{2}\D r\,,
\end{eqnarray}
where $\D r$ contains the full MSSM one-loop corrections \cite{deltarSUSY} and the leading two-loop QCD corrections \cite{deltar2loopaas}.
The leading contribution $\D\a\equiv\dot\Pi_T^{AA}(0)-{\rm Re}\Pi_T^{AA}(m_Z^2)/m_Z^2$ in $\D r$ cancels the uncertainty 
in the transverse photon self-energy. Both schemes agree within a few per mill in the final results. 

\subsubsection{Higgs sector}
In the MSSM two complex Higgs doublets $H_1$ and $H_2$ are required. The ratio of the vacuum expectation values is defined by
\begin{equation}
\tan\b \equiv \frac{v_2}{v_1}\,.
\end{equation}
The renormalization of this mixing angle is by no means trivial \cite{tanbgauge}.
In a definition by a specific physical process technical difficulties are introduced.
A process independent renormalization on the other hand leads to gauge dependence and/or numerical instabilities. 
A convenient choice, is the condition, that the pseudo-scalar Higgs field $A^0$ does not mix with the $Z$ vector boson for
on-shell momenta \cite{pokorski, dabelstein}.
\begin{eqnarray}
\frac{\d\tan\b}{\tan\b} &=& \frac{1}{m_Z\,\sin 2\b}\,\,\Im\!\left[\ReTilde\,\Pi_{A^0Z}(m_{A^0}^2)\right]\,.
\end{eqnarray}
However, the result is gauge dependent and can lead to big corrections for large $\tan\b$.
It can be improved by taking only the divergent part of the $A^0Z$ self energy, which makes
it numerically more stable and in the class of $R_\xi$ gauges $\xi$ independent at one-loop level. 
We call the first scale independent definition ``on-shell'' in contrast to the second ``$\drbar$ renormalization'' at a certain scale $Q$.
Both definitions and the translation between them are used in this work.

\subsubsection{SUSY sector}
For the renormalization of the chargino, neutralino and sfermion sector, we closely follow ref.\cite{mcvienna}.
We slightly extend the scheme for the general MSSM with complex phases.
For other on-shell renormalization procedures, see e.g.\cite{mchollik}.
After rotation from the interaction to the mass eigenstate basis, we introduce again wave function and mass counter terms
\begin{eqnarray}
&\nch_i \rightarrow (\d_{ij}+\onehf\d\tilde Z^L_{ij}P_L+\onehf\d\tilde Z^R_{ij}P_R)\nch_j\,,\quad
m_{\nch_i}\rightarrow m_{\nch_i} + \d m_{\nch_i}\,,&\\
&\sf\rightarrow (1+\onehf\d Z^\sf) \sf\,,\quad m^2_{\sf}\rightarrow m^2_{\sf} + \d m^2_{\sf}\,,\quad
{\rm for } \,\,\, \sf=\tilde{e}_L,\tilde{e}_R,\tilde{\nu}_e\,,&
\end{eqnarray}
where $\nch$ stands for both, charginos or neutralinos.
These counter terms are related to the corresponding self-energies by
\begin{eqnarray}
\d m_{\nch_i} &=& \frac{1}{2} \ReTilde\left[m_{\nch_i}\left(\Pi_{ii}^L(m^2_{\nch_i})+\Pi_{ii}^R(m^2_{\nch_i})\right)+
\Pi_{ii}^{SL}(m^2_{\nch_i})+\Pi_{ii}^{SR}(m^2_{\nch_i})\right],\\
\d\tilde Z^L_{ii} ~&=&~ \ReTilde \bigg[
-\Pi_{ii}^{L} (m_{\nch_i}^2)- m_{\nch_i}^2(\dot\Pi_{ii}^{L} (m_{\nch_i}^2)+ \dot\Pi_{ii}^{R}
(m_{\nch_i}^2))+\frac{1}{2 m_{\nch_i}}(\Pi_{ii}^{SL} (m_{\nch_i}^2)-\Pi_{ii}^{SR}
(m_{\nch_i}^2))\nonumber\\ && 
\hspace{2cm} - m_{\nch_i}(\dot\Pi_{ii}^{SL}
(m_{\nch_i}^2)+\dot\Pi_{ii}^{SR} (m_{\nch_i}^2))\bigg],\\
\d\tilde Z^L_{ij}~&=&~c_{ij}\, \ReTilde\left[
m_{\nch_j}^2\Pi_{ij}^{L} (m_{\nch_j}^2)+m_{\nch_i}m_{\nch_j}\Pi_{ij}^{R}(m_{\nch_j}^2)
+m_{\nch_i}\Pi_{ij}^{SL}(m_{\nch_j}^2)+m_{\nch_j}\Pi_{ij}^{SR}(m_{\nch_j}^2)\right],
\non \hspace{-20mm}\\ 
\end{eqnarray}
\begin{eqnarray}
\d\tilde Z^R_{ii}~=~\d\tilde Z^L_{ii}(L \leftrightarrow R)\,,\quad
\d\tilde Z^R_{ij}~=~\d\tilde Z^L_{ij}(L \leftrightarrow R)\,,\quad
c_{ij} = 2/(m_{\nch_i}^2-m_{\nch_j}^2)\,,
\end{eqnarray}
with $\d\tilde Z\equiv\d\tilde Z^\pm/\d\tilde Z^0$ and $\Pi_{ii}(k^2)$ the chargino/neutralino self-energies, respectively.
The corresponding counter terms for the sfermions are given by
\begin{equation}
\d m^2_{\sf}=\ReTilde\, \Pi^\sf(m^2_{\sf})\,,\quad \d Z^\sf=-\ReTilde\dot\Pi^\sf(m^2_{\sf})\,,\qquad
{\rm for}\quad\sf=\tilde{e}_L,\tilde{e}_R,\tilde{\nu}_e\,.
\end{equation}
Furthermore we have to define the neutralino and chargino rotation matrices at one-loop level.
A process and scale independent fixing seems appropriate. We define the counter terms in such a way
that they cancel the rotation, induced by the antihermitian parts of the off-diagonal wave function corrections.
\begin{eqnarray}\label{nrotcts}
&\d N_{ij}=\displaystyle{\frac{1}{4}\sum_{k=1}^4}\left(\d\tilde Z^{0,L}_{ik}-\d\tilde Z^{0,R}_{ki}\right)N_{kj}\,,&\\\label{crotcts}
&\d U_{ij}=\displaystyle{\frac{1}{4}\sum_{k=1}^4}\left(\d\tilde Z^{\pm,R}_{ik}-(\d\tilde Z^{\pm,R}_{ki})^*\right)U_{kj}\,,\quad
\d V_{ij}=\displaystyle{\frac{1}{4}\sum_{k=1}^4}\left(\d\tilde Z^{\pm,L}_{ik}-(\d\tilde Z^{\pm,L}_{ki})^*\right)V_{kj}\,.&
\end{eqnarray}
In \cite{Yamada} it is shown that this fixing eqs.~(\ref{nrotcts}) and (\ref{crotcts}), calculated within the Feynman-t'Hooft gauge, 
can be regarded as a gauge independent one. 
The counter terms for the on-shell masses and the rotation matrices directly yield the counter terms for the mass matrix elements
\begin{eqnarray}\label{nmmcts}
\d Y_{ij}&=&\frac{1}{2} \sum_{l,n=1}^4 \ZN_{ni}\ZN_{lj}\,\ReTilde \left[ m_{\cch_n}
\Pi^L_{nl} (m_{\cch_n}^2) + m_{\cch_l}
\Pi^R_{nl}(m_{\cch_l}^2)+\Pi^{SR}_{nl}(m_{\cch_l}^2)+\Pi^{SL}_{nl}(m_{\cch_n}^2)
\right]\,,\non\\ \\ \label{cmmcts}
\d X_{ij}&=&\frac{1}{2} \sum_{l,n=1}^4 U_{ni} V_{lj}\,\ReTilde \left[ m_{\cchpm_n}
\Pi^L_{nl} (m_{\cchpm_n}^2) + m_{\cchpm_l}
\Pi^R_{nl}(m_{\cchpm_l}^2)+\Pi^{SR}_{nl}(m_{\cchpm_l}^2)+\Pi^{SL}_{nl}(m_{\cchpm_n}^2)
\right]\,,\non\\
\end{eqnarray}
Note that in the presence of complex mass parameters this renormalization automatically includes the one-loop definition of
CP violating phases.\\
For a proper one-loop calculation each involved Lagrangian parameter requires a clear definition, i.e. a unique counter term.
Since not all entries in the neutralino and chargino mass matrices are independent parameters, in general the counter terms eqs.~(\ref{nmmcts})
and (\ref{cmmcts})
cannot be interpreted as the counter terms to the Lagrangian parameters given in the tree-level mass matrices.
For example the parameter $M$ cannot be defined by the counter term $\d Y_{22}$ to the neutralino mass matrix and in the same computation
by the chargino mass matrix counter term $\d X_{11}$. The same holds holds for $\mu$ and the parameters already fixed in the gauge
and Higgs sector. Therefore, we define $M$ and $\mu$ to be the parameters in the chargino mass matrix $M\equiv X_{11}$, $\mu\equiv X_{22}$
and further $M'\equiv Y_{11}$ throughout the calculation. For all other elements finite shifts have to be taken into account, e.g. 
$Y_{22}=M+\d X_{11}-\d Y_{22} = M+ \D M$. The UV finiteness of these shifts is a nontrivial check of this method. 
The so obtained one-loop corrected mass matrices give after diagonalization the one-loop on-shell neutralino and chargino masses.
The corresponding rotation matrices have the appropriate values for the counter terms eqs.~(\ref{nrotcts}) and (\ref{crotcts}).
This procedure of defining on-shell parameters can be simply extended to the sfermion sector.
In our simplified case without left-right mixing, there are two free parameters for three masses. 
We fix $\Msl$ and $\Mse$ in such a way, that the selectron masses do not obtain one-loop corrections 
\begin{eqnarray}
\d \Msl^2 &=& \d m^2_{\tilde{e}_L}+\d\Big(m_Z^2 \cos2\b \left( \onehf - \sin^2\theta_W \right)\Big) 
\,,\\
\d \Mse^2 &=& \d m^2_{\tilde{e}_R}+\d\left(m_Z^2 \cos2\b \, \sin^2\theta_W\right) 
\,.
\end{eqnarray}
Due to a finite shift in $\Msl$, the sneutrino mass obtains the one-loop correction
\begin{eqnarray}
\D m^2_{\tilde{\nu}_e} &=& \d \Msl^2 + \frac{1}{2}\,\d\!\left(m_Z^2 \cos2\b\right) - \d m^2_{\tilde{\nu}_e}\,.
\end{eqnarray}

\section{QED corrections}\label{QED corrections}
The full one-loop corrections include diagrams with virtual photons attached to the tree-level diagrams.
These contributions are IR divergent and regularized by an infinitesimal photon mass $\la$. Concerning chargino
and neutralino production these diagrams cannot be separated from the residual weak corrections
in a gauge invariant and UV finite way. This can be traced back to the tree-level selectron and sneutrino t-channel diagrams,
which introduce the charged current coupling $g\equiv e/s_w$. The same effect can be observed
in the SM, e.g. for $W$ pair production \cite{Wpair}. 
The cross sections become IR finite and thus physically meaningful only by inclusion of real photon emission.
For the calculation of the real photonic corrections, we use the so-called phase-space slicing method~\cite{psslicing}.
The singular soft and collinear parts in the bremsstrahlung phase space are separated from the finite region. 
Both contributions can be written proportional to the tree-level cross-section, up to small terms of $O(\D E/\sqrt{s})$
and $O(\D\t)$, and performed analytically. 
The collinear (soft) singularities are regularized by the electron (infinitesimal photon) mass and cancel
the corresponding terms in the virtual corrections.
In the soft photon area only photons up to a CM energy $\D E$ are considered
\begin{eqnarray}\label{softphoton1}
\ds^{\rm soft} &=& -\ds^{\rm tree}\frac{\a}{4\pi^2}\int_{k_\g^0\leq\D E} \frac{d^3{\bf k_\g}}{k_\g^0}
\left(\frac{p_1^\mu}{p_1 k_\g}-\frac{p_2^\mu}{p_2 k_\g}+Q_\nch\frac{k_1^\mu}{k_1 k_\g}-Q_\nch\frac{k_2^\mu}{k_2 k_\g}\right)^2\,,
\end{eqnarray} 
whereas $Q_\nch=-1/0$ stands for the electric charge of the chargino/neutralino respectively. The result can be expressed in terms
of the soft photon integrals given in ref.~\cite{Denner}
\begin{eqnarray}\label{softphoton2}
&\ds^{\rm soft}\,=\,-\ds^{\rm tree}\displaystyle{\frac{\a}{4\pi^2}}\left(\d_{\rm soft}^{\rm ISR}+\d_{\rm soft}^{\rm FSR}+\d_{\rm soft}^{\rm ISR-FSR}\right)\,,&\\
&\d_{\rm soft}^{\rm ISR}\,=\,I_{p_1p_1}+I_{p_2p_2}-2I_{p_1p_2}\,,\quad \d_{\rm soft}^{\rm FSR}=Q_\nch^2\left(I_{k_1k_1}+I_{k_2k_2}-2I_{k_1k_2}\right)\,,&\\
&\d_{\rm soft}^{\rm ISR-FSR}\,=\,2Q_\nch\left(I_{p_1k_1}+I_{p_2k_2}-I_{p_1k_2}-I_{p_2k_1}\right)\,.&
\end{eqnarray} 
The collinear part takes hard photons in a small angle $\D\t$ around the beam axis into account
\begin{eqnarray}\label{collinear1}
\int\ds^{\rm coll}(p_1,p_2,\xi_-,\xi_+) &=& \frac{\a}{2\pi}\int^{1-2\D E/\sqrt{s}}_0 {\rm d}x\sum_{\a=\pm} G_\a(s,x,\D\t)\,.\nonumber\\
&&\hspace{-2cm} .\left[\int\ds^{\rm tree}(xp_1,p_2,\a\xi_-,\xi_+)+\int\ds^{\rm tree}(p_1,xp_2,\xi_-,\a\xi_+)\right]\,,
\end{eqnarray}
with
\begin{eqnarray}
G_+(s,x,\D\t)\,=\,\frac{1+x^2}{1-x}\left[\log\left(\frac{s\D\t^2}{4m_e^2}\right)-1\right]\,,\quad G_-(s,x,\D\t)\,=\,1-x\,.
\end{eqnarray} 
The finite hard bremsstrahlung has to be calculated by integration of the squared tree-level matrix-element for
$e^+e^- \rightarrow \nch_i\nch_j\g$ over the three-particle final-state phase-space.
The complete $O(\a)$ corrections can then be written as sum of virtual, soft, collinear, and finite contributions,
all depending on unphysical auxiliary parameters.
\begin{eqnarray}
\D\s_{O(\a)} = \int\!\left(\ds^{\rm virt}(\la) + \ds^{\rm soft}(\la,\D E)\right) + \int\!\ds^{\rm coll}(\D E, \D\t) + 
\int\!\ds^{\rm finite}(\D E, \D\t)
\end{eqnarray} 
Summing up the contributions in $\D\s_{O(\a)}$, we obtain a cut-off independent result.
This has been checked analytically for $\la$ and numerically for $\D E$ and $\D\t$ in the intervals 
$10^{-5}\leq\D E/\sqrt{s}\leq 10^{-2}$ and $10^{-3}\leq\D\t\leq 10^{-2}$.\\
For precise predictions of chargino and neutralino pair production higher orders beyond $O(\a)$ have to be taken into account.
The structure function formalism \cite{sff} provides the possibility of defining process-independent logarithmic QED corrections, 
which originate from collinear virtual and real photons radiated off the incoming electron-positron beams.
\begin{eqnarray}\label{univ} 
\int\!\ds^{\rm tree}+\int\!\ds^{\rm univ} =
\int_0^1\! dx_1\!\int_0^1\!dx_2\,\G_{ee}^{\rm LL}(x_1,Q^2) \G_{ee}^{\rm LL}(x_2,Q^2)\,\int\!\!\ds^{\rm tree}(x_1p_{e^-},x_2p_{e^+})\,.
\end{eqnarray} 
We use the leading-log structure function up to $O(\a^3)$, given in ref.\cite{Sk90}.
\begin{eqnarray}\label{GLL}
\G^{\rm LL}_{ee}(x,Q^2) &=& \frac{\exp(-\frac{1}{2}\b\,\g_E+\frac{3}{8}\b)}{\G(1+\frac{\b}{2})}\frac{\b}{2}(1-x)^{\frac{\b}{2}-1}\nonumber\\
&&\hspace{-2cm}-\frac{\b}{4}(1+x)+\frac{\b^2}{16}\Big(-2(1+x)\log(1-x)-\frac{2\log x}{1-x}+\frac{3}{2}(1+x)\log x-\frac{x}{2}-\frac{5}{2}\Big)\nonumber\\
&&\hspace{-2cm}+(\frac{\b}{2})^3\left[-\frac{1}{2}(1+x)\left(\frac{9}{32}-\frac{\pi^2}{12}+\frac{3}{4}\log(1-x)+\frac{1}{2}\log^2(1-x)-\frac{1}{4}\log x\log(1-x)
\right. \right.\nonumber\\
&&\hspace{-2cm}\left. \left.+\frac{1}{16}\log^2x-\frac{1}{4}{\rm Li}_2(1-x)\right)+\frac{1}{2}\frac{1+x^2}{1-x}\Big(-\frac{3}{8}\log x+\frac{1}{12}\log^2x
-\frac{1}{2}\log x\log(1-x)\Big)\right.\nonumber\\
&&\hspace{-2cm}\left.-\frac{1}{4}(1-x)\Big(\log(1-x)+\frac{1}{4}\Big)+\frac{1}{32}(5-3x)\log x\right]\,,
\end{eqnarray}  
with the gamma function $\G$, the Euler constant $\g_E\sim 0.577216$, and \mbox{$\b=\frac{2\a}{\pi}(\log\frac{Q^2}{m_e^2}-1)$}.
For simplicity we fix the free scale $Q^2=s$ throughout the calculations.
Since $\b$ contains in addition to the $\log\frac{Q^2}{m_e^2}$ a constant term also non-leading log terms are included to form the correct 
ISR soft-photon pole, see also eq.~(\ref{defweak}).
Furthermore, the soft-photon contributions are resummed up to all orders in perturbation theory.\\
Since the above calculated correction  $\D\s_{O(\a)}$ already contains the universal terms of $O(\a)$
\begin{eqnarray}\label{univ1} 
\int\!\!\ds^{\rm univ,1} &=& 
\int_0^1\! dx_1\!\int_0^1\!dx_2\,\G_{ee}^{\rm LL,1}(x_1,Q^2) \G_{ee}^{\rm LL,1}(x_2,Q^2)\,\int\!\!\ds^{\rm tree}(x_1p_{e^-},x_2p_{e^+})\,,
\end{eqnarray} 
with
\begin{eqnarray}\label{GLL1}
\G_{ee}^{\rm LL,1}(x,Q^2)~&=&~\frac{\b}{4}\lim_{\e\rightarrow 0}\{\d(1-x)[\frac{3}{2}+2\log(\e)] +\theta(1-x-\e)\frac{1+x^2}{1-x}\}\,,
\end{eqnarray}
we have to subtract them from the complete corrections to avoid double counting.
\begin{eqnarray}
\int\!\ds^{\rm complete} &=& \int\!\ds^{\rm tree}+\D\s_{O(\a)}+\int\!\left(\ds^{\rm univ}-\ds^{\rm univ,1}\right)
\end{eqnarray}

\subsection*{Definition of weak and QED corrections}
In spite of the impossibility to disentangle the different contributions to the $O(\a)$ corrections, 
it is of special interest to distinguish the genuine weak corrections
from the large and on experimental cuts dependent photon part.
In fact, there is no unique way of doing this.
A naive treatment would be to take the pure virtual corrections and set the photon mass equal to a typical scale 
of the corresponding process $\la\equiv Q$.
However, this leaves us with enhanced Sudakov double-logarithms $\log^2\frac{s}{m_e^2}$ from virtual soft photons attached
to the incoming beams, which are cancelled by the corresponding real, soft part.  
In the following we take the sum of virtual and soft corrections and extract the $\D E$ dependent terms as well as the
contributions proportional to $L_e\equiv\log\frac{s}{m_e^2}$, stemming from the collinear virtual+soft photons, eq.~(\ref{univ1})
\protect\footnote{The definitions in the previous work~\cite{NeuProdPaper} slightly differ from those given here.}.
\begin{eqnarray}\label{defweak}
\ds^{\rm weak} &=& \ds^{\rm virt+soft}(\D E)-\frac{\a}{\pi}\ds^{\rm tree}\left[\log{\frac{4\D E^2}{s}}(L_e-1+\D_{\g})+ 
\frac{3}{2}L_e\right]\,.
\end{eqnarray}
The term $\D_{\g}$ takes the cut-off dependent terms from final state radiation (FSR) and ISR-FSR interference of eq.~(\ref{softphoton2}) 
into account.
The sum $\ds^{\rm tree}+\ds^{\rm weak}$ is identical to the ``reduced genuine SUSY cross section'' within the SPA convention \cite{SPA}.
The full corrected cross-section can now be obtained by the sum
\begin{eqnarray}
\int\!\ds^{\rm complete} &=& \int\!\left(\ds^{\rm tree}+\ds^{\rm weak}\right)+\int\!\ds^{\rm non-univ}+\int\!\ds^{\rm univ}\,,
\end{eqnarray}
with the non-universal QED corrections
\begin{eqnarray}
\int\!\!\ds^{\rm non-univ} &=& \int\!\!\ds^{\rm coll+finite}(\D E)-\int\!\ds^{\rm univ,1}+\nonumber\\
&&\hspace{3cm} \frac{\a}{\pi}\int\!\!\ds^{\rm tree}\left[\log{\frac{4\D E^2}{s}}(L_e-1+\D_{\g})+ \frac{3}{2}L_e\right]\,.
\end{eqnarray}
A second way to highlight weak corrections is to compare the complete corrected cross section
(including the hard photon radiation) with an improved tree-level $\ds^{\rm tree+ISR}$ that already contains the universal
QED corrections. 
\begin{eqnarray}
& \int\!\ds^{\rm complete} = \int\!\ds^{\rm tree+ISR}+\int\!\ds^{\rm residual}\,, &\\
& \int\!\ds^{\rm tree+ISR} = \int\!\ds^{\rm tree} + \int\!\ds^{\rm univ}\,,\qquad
\int\!\ds^{\rm residual} = \int\!\ds^{\rm weak} + \int\!\ds^{\rm non-univ}\,.&
\end{eqnarray} 
The advantage of this definition is that it does not require a more or less superficial splitting of virtual and real corrections.
On the other hand, the ``residual'' corrections include the non-universal QED corrections. They are in general small, 
but can be comparable to the loop corrections, especially the ISR-FSR terms.
Furthermore, it can be inconvenient for technical reasons to include the hard bremsstrahlung process in the definition of a ``weak'' correction. 

\section{Numerical results}\label{Numerical results}
\begin{figure}[h!]
\begin{center}
\mbox{\mbox{\resizebox{84mm}{!}{\includegraphics{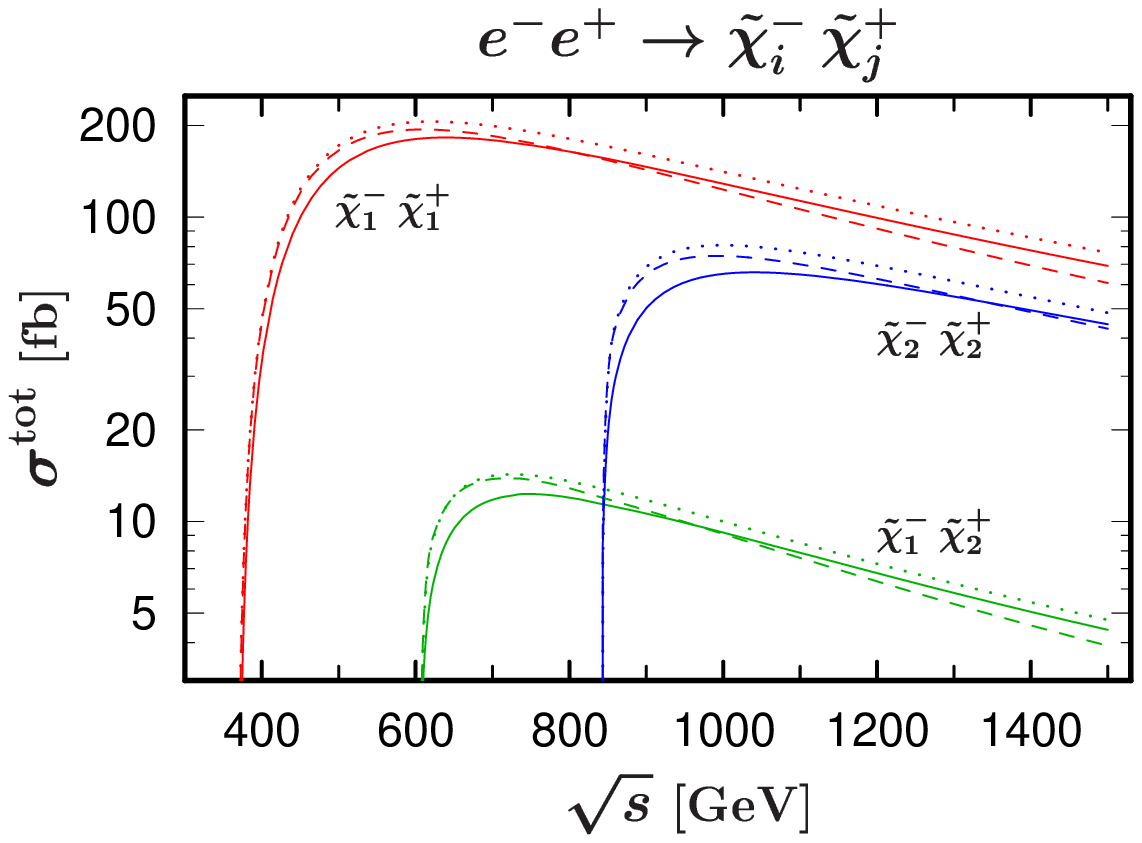}}}\hspace{-5mm}
\mbox{\resizebox{84mm}{!}{\includegraphics{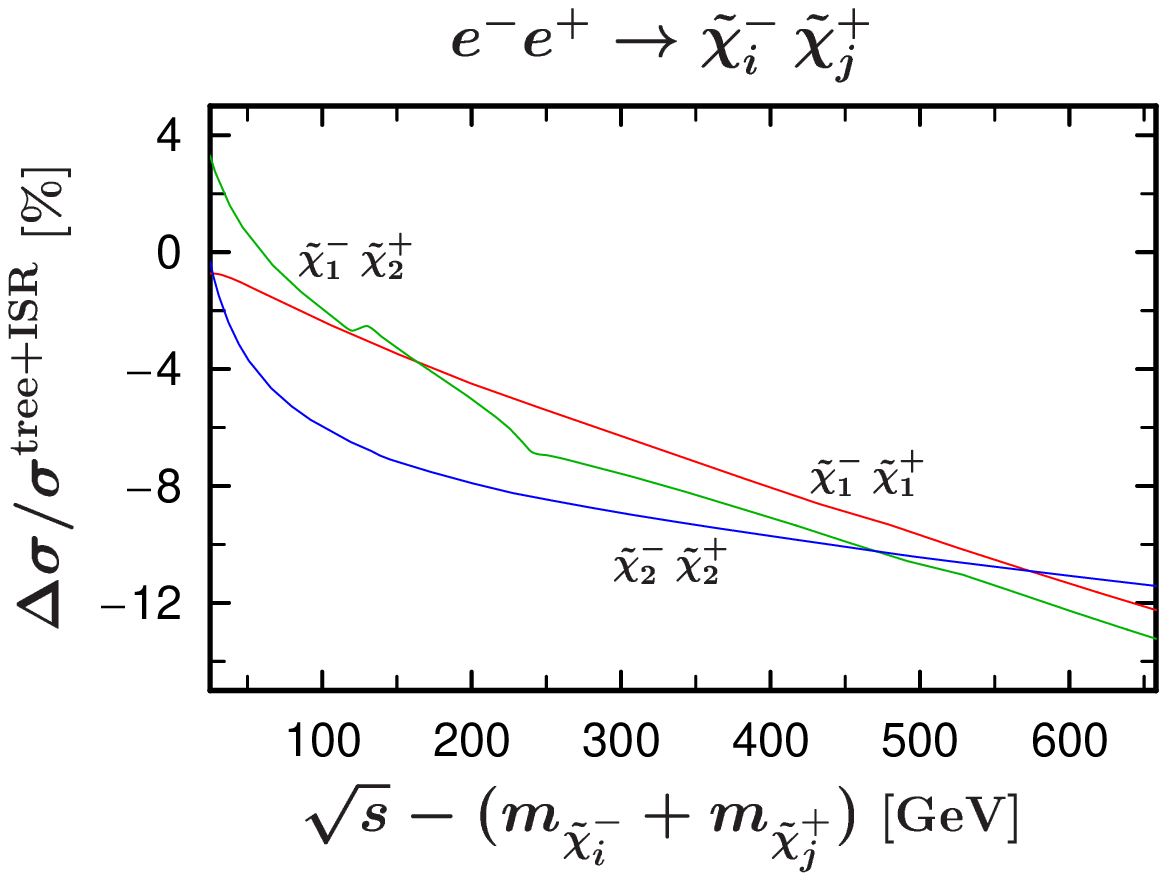}}}}
\end{center}
\caption{Left: Total cross sections for chargino pair production at tree-level \{dotted\}, with weak \{dashed\}, and with complete corrections \{solid\}.
Right: Complete corrections relative to the improved tree-level above the particular production threshold.}
\label{fig:cha_tot1}
\end{figure}

\begin{figure}[h!]
\begin{center}
\hspace*{-1mm}
\mbox{\mbox{\resizebox{84mm}{!}{\includegraphics{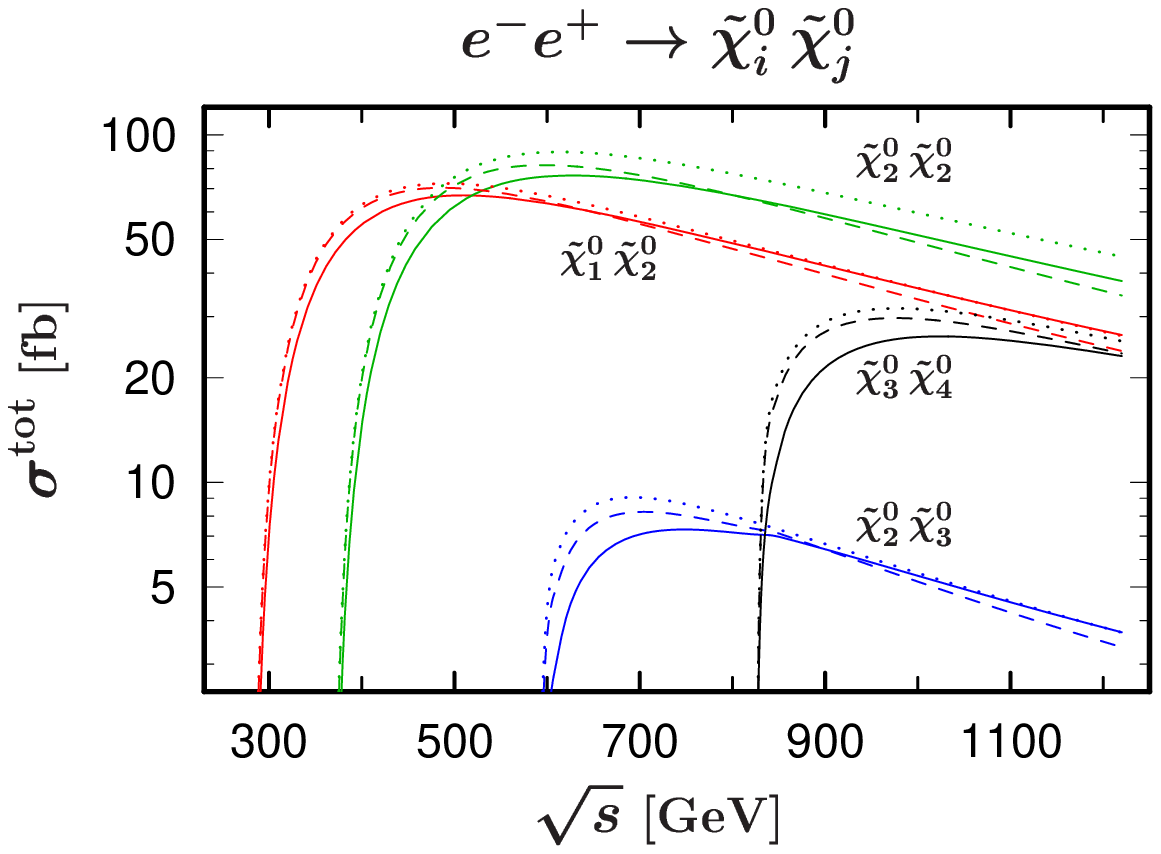}}} \hspace{-5mm}
\mbox{\resizebox{84mm}{!}{\includegraphics{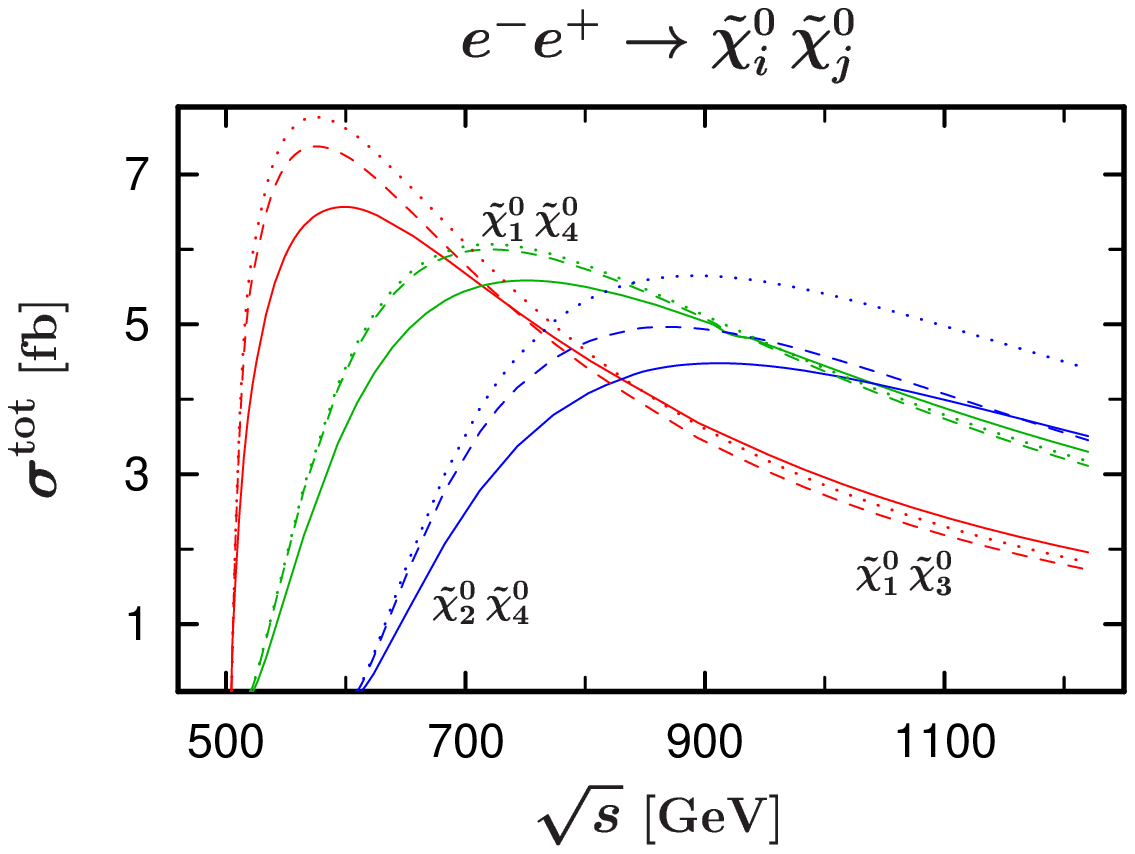}}}}
\caption{Total cross sections for neutralino pair production at tree-level \{dotted\}, with weak \{dashed\}, and with complete corrections \{solid\}.}
\label{fig:neu_tot12}
\end{center}
\end{figure}

For the numerical analysis we concentrate especially on the SPS1a' point, proposed in the SPA project \cite{SPA}.
It is close to the original Snowmass SPS1a scenario and compatible with all available precision data and actual
mass and cosmological bounds.
The parameters in the SPA convention are defined in the \drbar scheme at the scale of $Q$ = 1 TeV.
A translation from these parameters to our on-shell definition can be simply performed by subtraction of the corresponding
counter terms, i.e. ${\cal P}^{\rm OS}={\cal P}(Q)-\d{\cal P}(Q)$. The used values in this work can be
seen in Table \ref{tab:SPS1ap}.
\begin{table}
\begin{center}
\begin{tabular}{|c||c|c|}
   \hline
  ${\cal P}$ & \drbar & {\rm OS} \\ \hline \hline
  $M'$ & 103.22 & 100.32\\
  $M$ & 193.31 & 197.03\\
  $\mu$ & 402.87 & 399.94\\
  $\tan\b$ & 10 & 10.31\\
  $\Mse$ & 115.59 & 117.71\\
  $\Msl$ & 181.25 & 183.98\\ \hline
\end{tabular}
\hspace{1cm}
\begin{tabular}{|c||c|c||c|}
   \hline
  ${\cal M}$ & {\rm OS} & ${\cal M}$ & {\rm OS}\\ \hline \hline
  $m_{\tilde\chi_1^\pm}$ & 184.2 & $m_{\cch_1}$ & 97.75\\
  $m_{\tilde\chi_2^\pm}$ & 421.1 & $m_{\cch_2}$ & 184.4\\
  $m_{\tilde{e}_L}$ & 190.1 & $m_{\cch_3}$ & 406.9\\
  $m_{\tilde{e}_R}$ & 125.2 & $m_{\cch_4}$ & 419.5\\
  $m_{\tilde{\nu}_e}$ & 172.8 & &\\ \hline
\end{tabular}
\end{center}
\caption[SPS1ap]{Parameters of the SPS1a' scenario in the \drbar and on-shell scheme and particle masses.}\label{tab:SPS1ap}
\end{table}
For all other parameters that are free of renormalization conditions, the \drbar or on-shell values can be used.
The difference is of higher order for the current processes.\\
Since we use an on-shell renormalization, the appropriate tree-level for the one-loop calculation is given in terms
of on-shell parameters. On the other hand, our original input are the $\drbar$ parameters of the SPS1a' scenario.
We therefore show the corrections relative to the tree-level in the SPA convention, i.e. calculated in terms of on-shell
masses and $\drbar$ values for all couplings. Using this tree-level definition, the relative corrections
are the same compared with those calculated in other renormalization schemes, up to terms of higher order.
For this purpose the used $\drbar$ value at $Q$=1 TeV of the fine structure constant is $\a=1/124.997$.
In the presented numerics, we use for the charge renormalization the $\a^{\rm eff}(m_{\scriptscriptstyle Z})|_{\overline{\rm MS}}$
scheme for neutralino production and the $G_F$ scheme for chargino production, as discussed in section~\ref{Gauge sector}. 
By calculating the same cross section in both schemes, we find good agreement within a few per-mill in the final results.
The Figs.~(\ref{fig:cha_tot1}) and (\ref{fig:neu_tot12}) show the total cross sections for chargino and neutralino pair production
at the tree-level in the SPA convention, and with weak and full corrections. Additionally, Fig.~(\ref{fig:cha_tot1})
shows the complete corrections to the improved tree-level, where ISR is already included. Since the soft-photon pole is already
absorbed into the tree-level, we have moderate corrections even near the threshold.
\begin{figure}[h!]
\begin{center}
\hspace*{-1mm}
\mbox{\mbox{\resizebox{84mm}{!}{\includegraphics{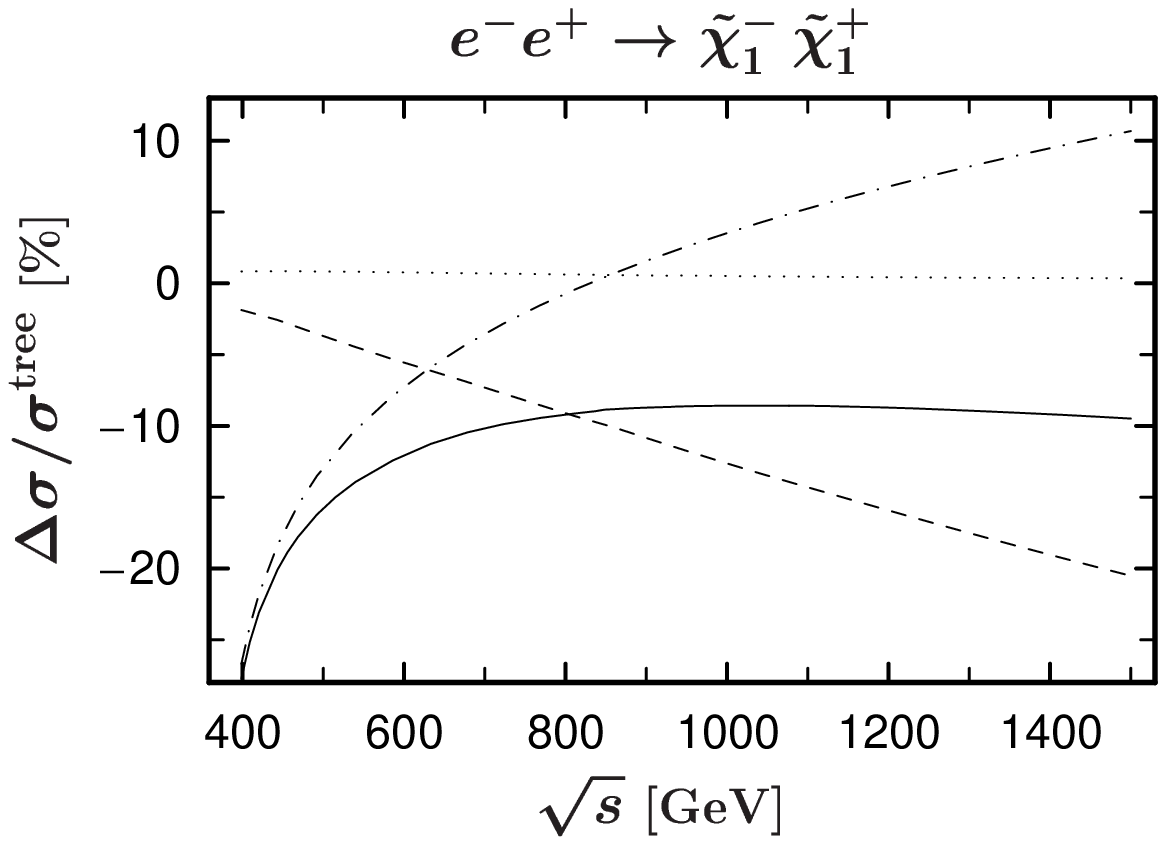}}} \hspace{-5mm}
\mbox{\resizebox{84mm}{!}{\includegraphics{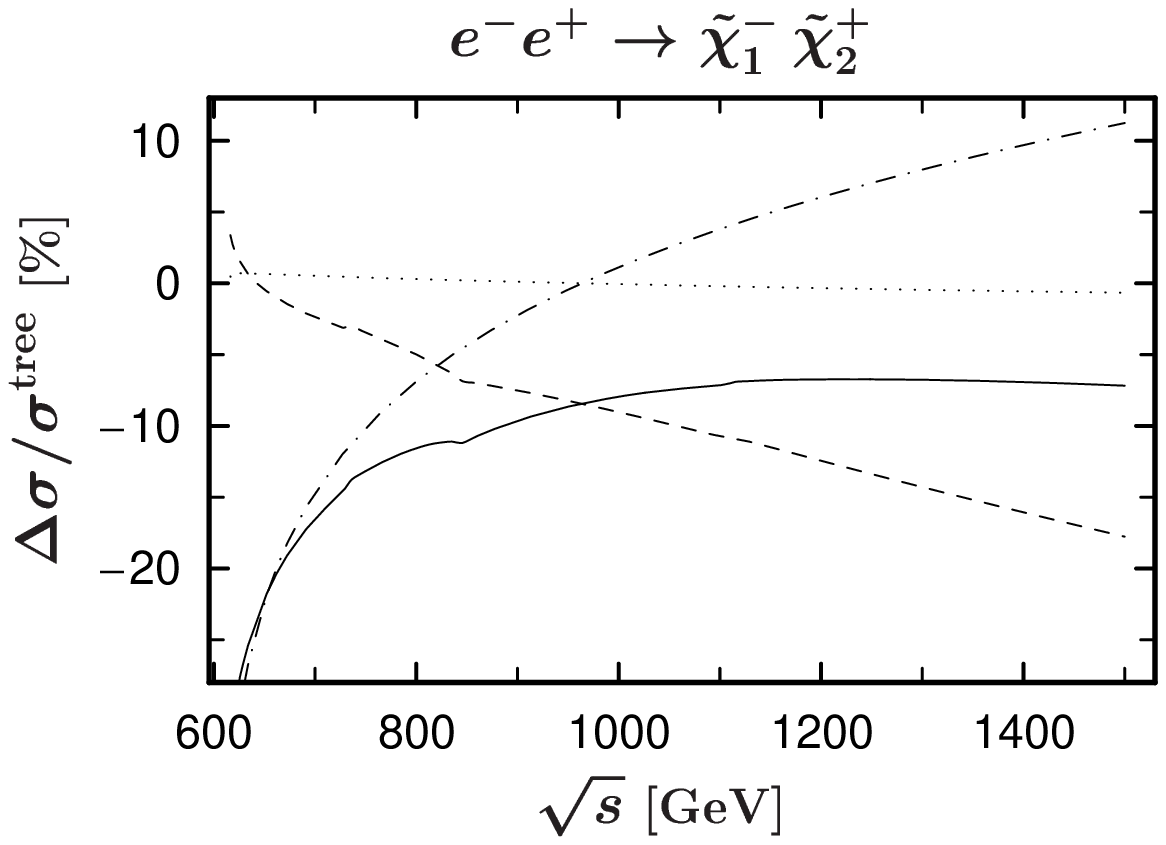}}}}
\caption{Radiative corrections for chargino pair production.
The \{full, dashed, dotted, dash-dotted\} line corresponds to the \{complete, weak, non-universal QED, universal QED\} corrections to the total
tree-level cross section.}
\label{fig:cha_corr12}
\end{center}
\end{figure}
\begin{figure}[h!]
\begin{center}
\hspace*{-1mm}
\mbox{\mbox{\resizebox{84mm}{!}{\includegraphics{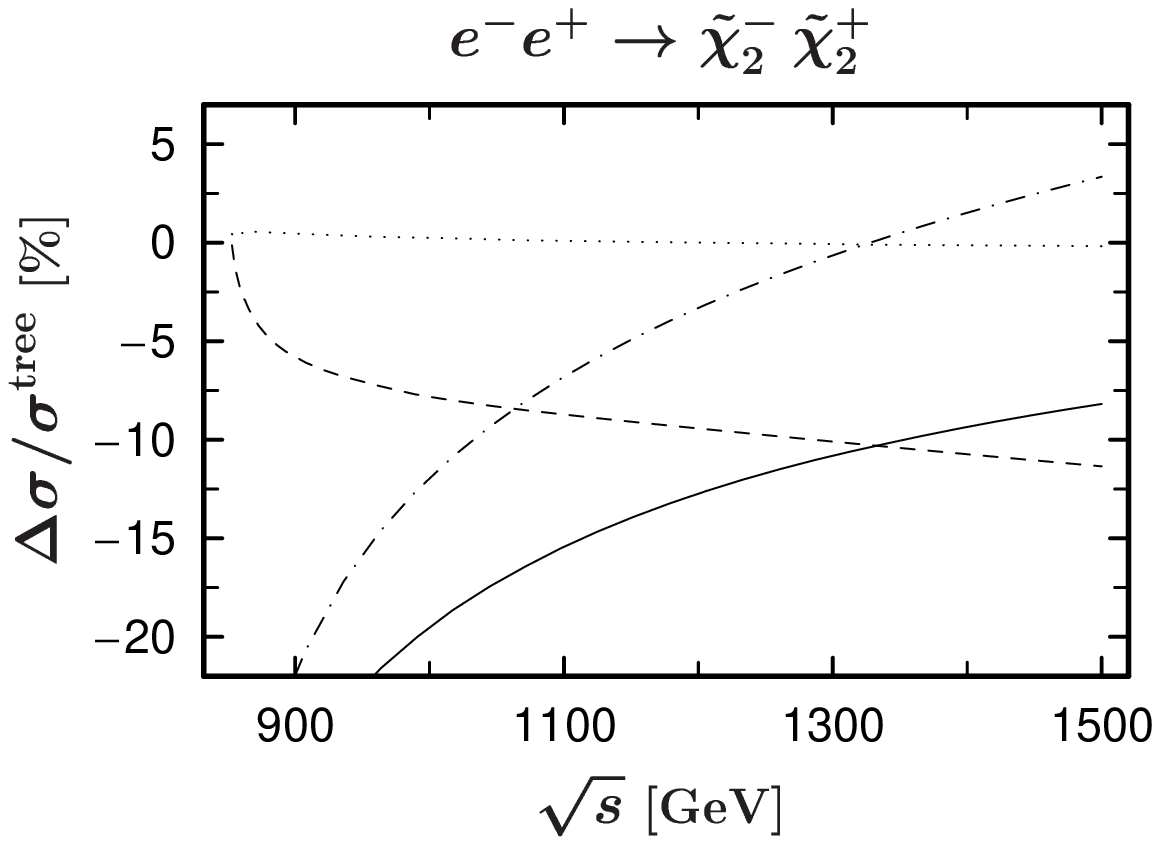}}} \hspace{-5mm}
\mbox{\resizebox{84mm}{!}{\includegraphics{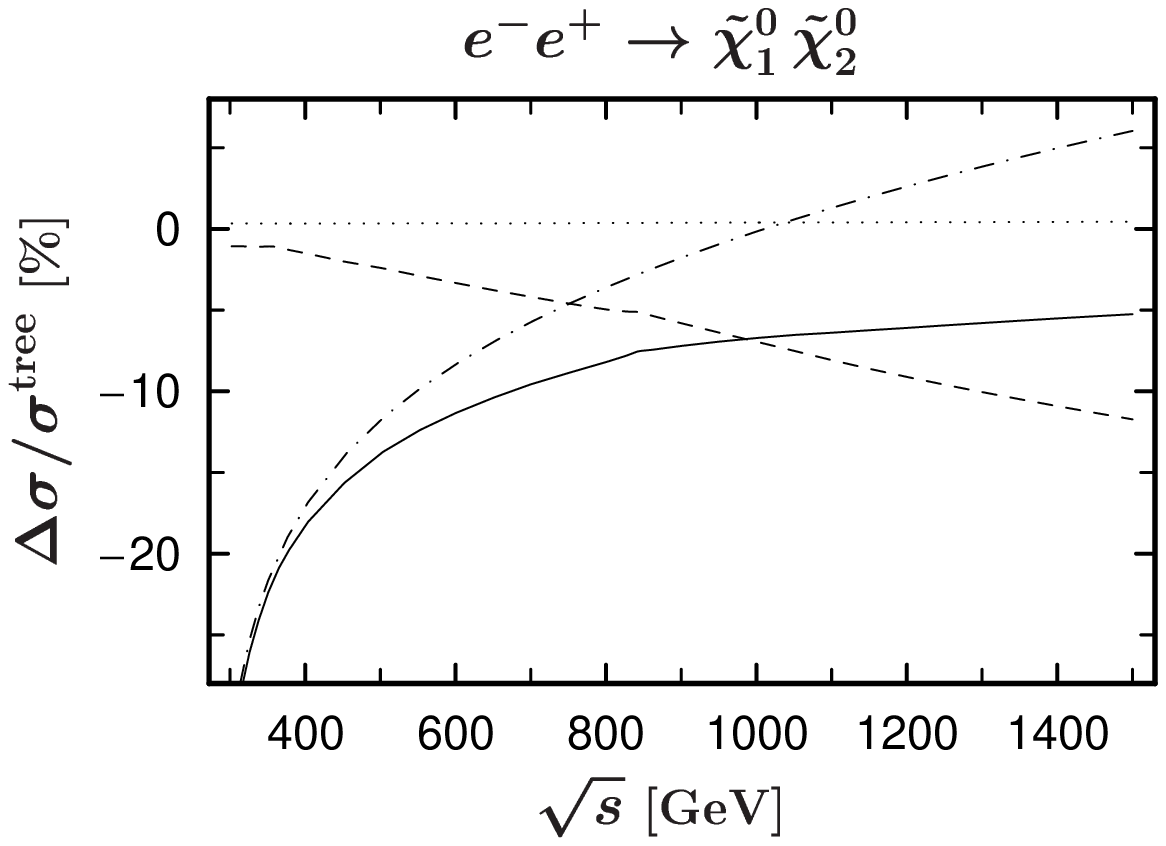}}}}
\caption{Radiative corrections for chargino and neutralino pair production.
The \{full, dashed, dotted, dash-dotted\} line corresponds to the \{complete, weak, non-universal QED, universal QED\} corrections to the total
tree-level cross section.}
\label{fig:cha_corr34}
\end{center}
\end{figure}
\begin{figure}[h!]
\begin{center}
\hspace*{-1mm}
\mbox{\mbox{\resizebox{84mm}{!}{\includegraphics{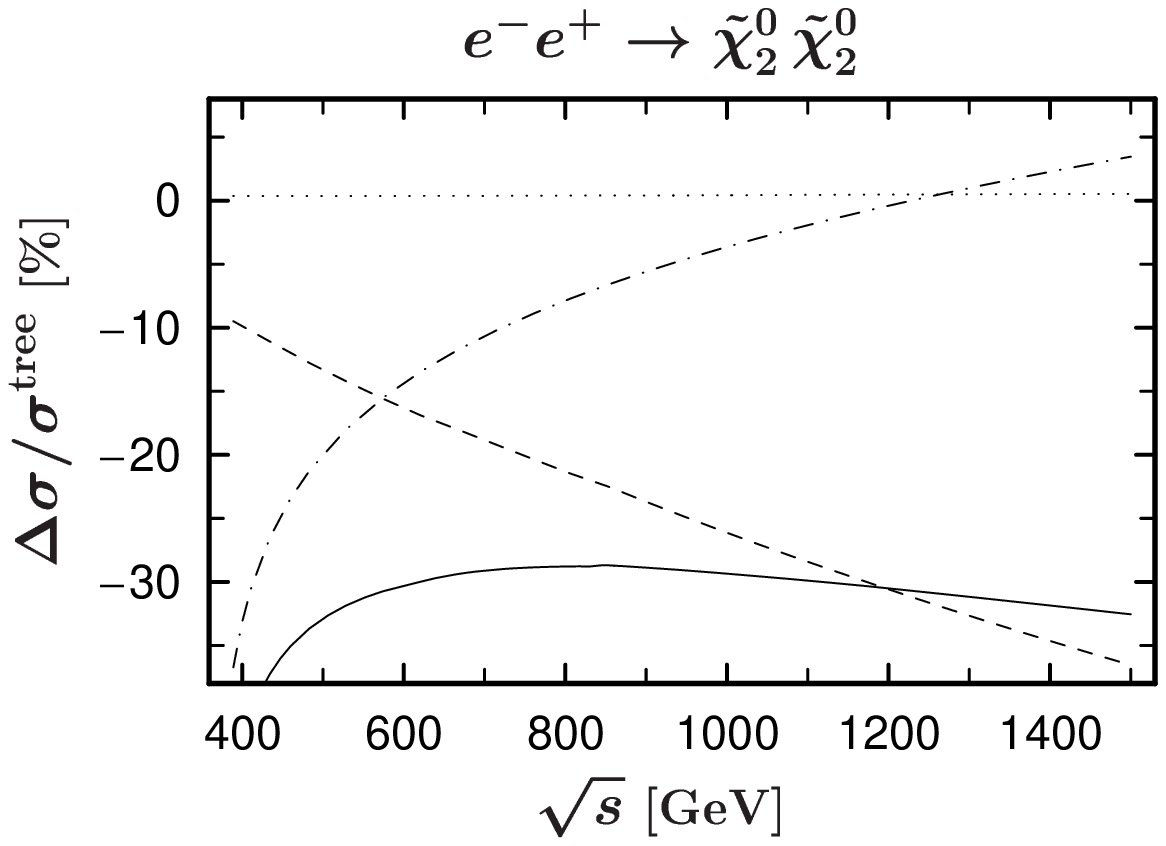}}} \hspace{-5mm}
\mbox{\resizebox{84mm}{!}{\includegraphics{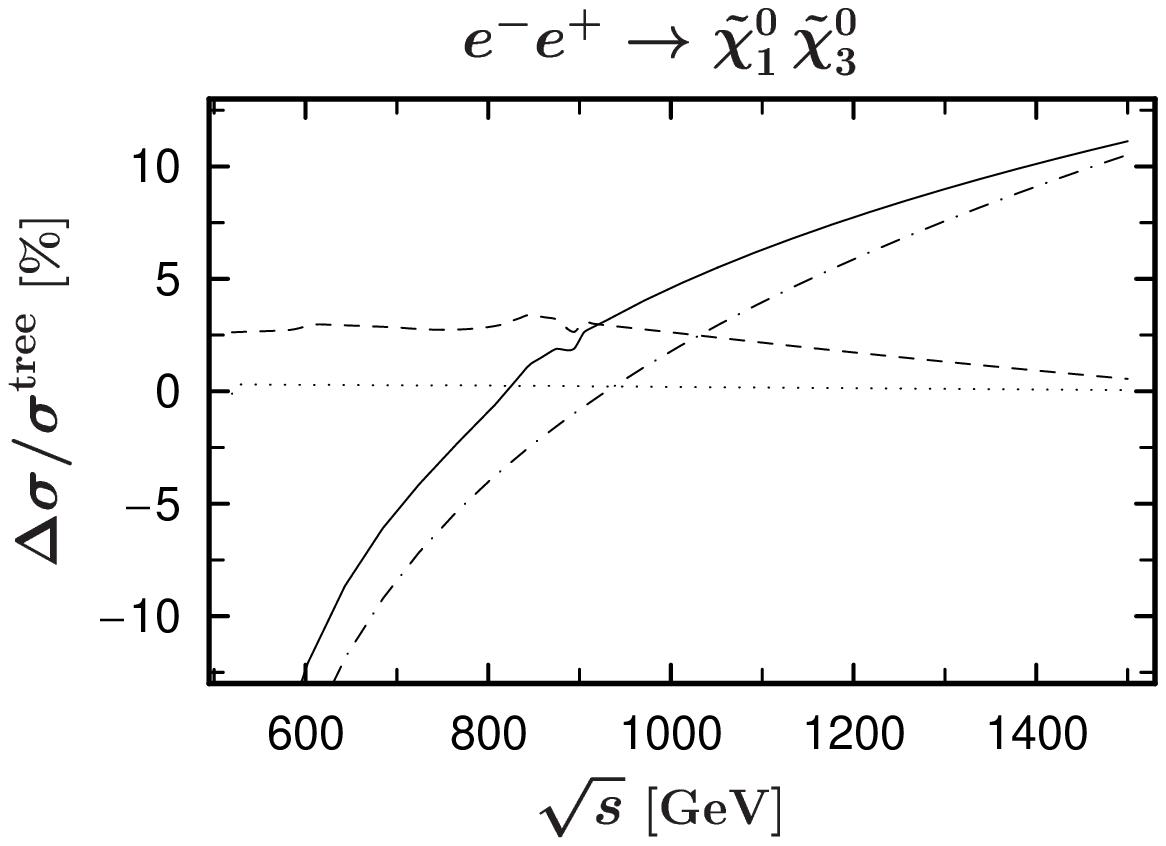}}}}
\caption{Radiative corrections for neutralino pair production.
The \{full, dashed, dotted, dash-dotted\} line corresponds to the \{complete, weak, non-universal QED, universal QED\} corrections to the total
tree-level cross section.}
\label{fig:neu_corr12}
\end{center}
\end{figure}

Comparing the various corrections to the total tree-level cross section for the different production channels
Figs.~(\ref{fig:cha_corr12}, \ref{fig:cha_corr34}, \ref{fig:neu_corr12}) some common characteristics can be
recognized. Near the threshold the negative soft photon contributions dominate. Far away from the threshold, the positive universal QED corrections
partially cancel the large and negative weak contributions. The almost constant non-universal QED corrections are small and in comparison 
with other corrections often negligible.
Due to the marginal non-universal QED corrections, the differences between the two proposed ways to highlight ''genuine weak" corrections are 
quite small. However, this does not have to be true any longer if cuts on the phase space are applied or distributions in particle energies
or scattering angles are discussed.
\begin{figure}[h!]
\begin{center}
\mbox{\mbox{\resizebox{84mm}{!}{\includegraphics{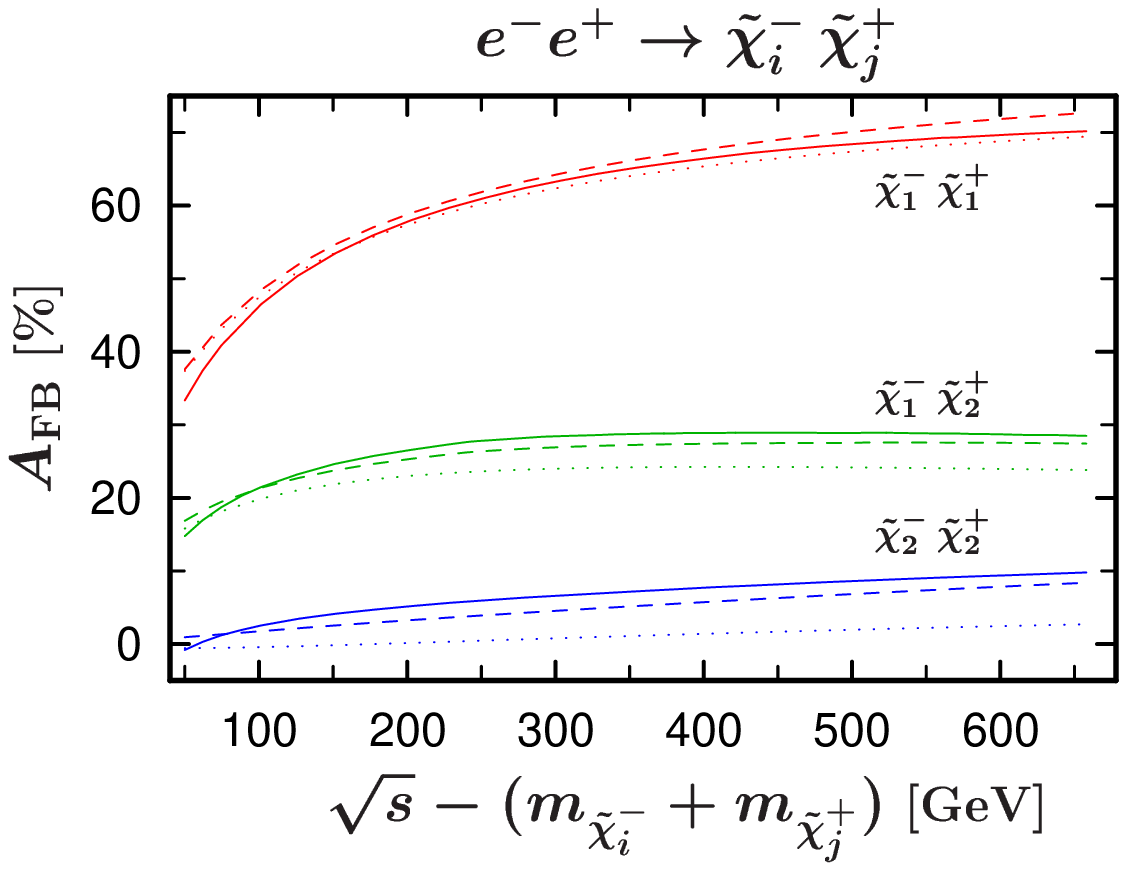}}}\hspace{-5mm}
\mbox{\resizebox{84mm}{!}{\includegraphics{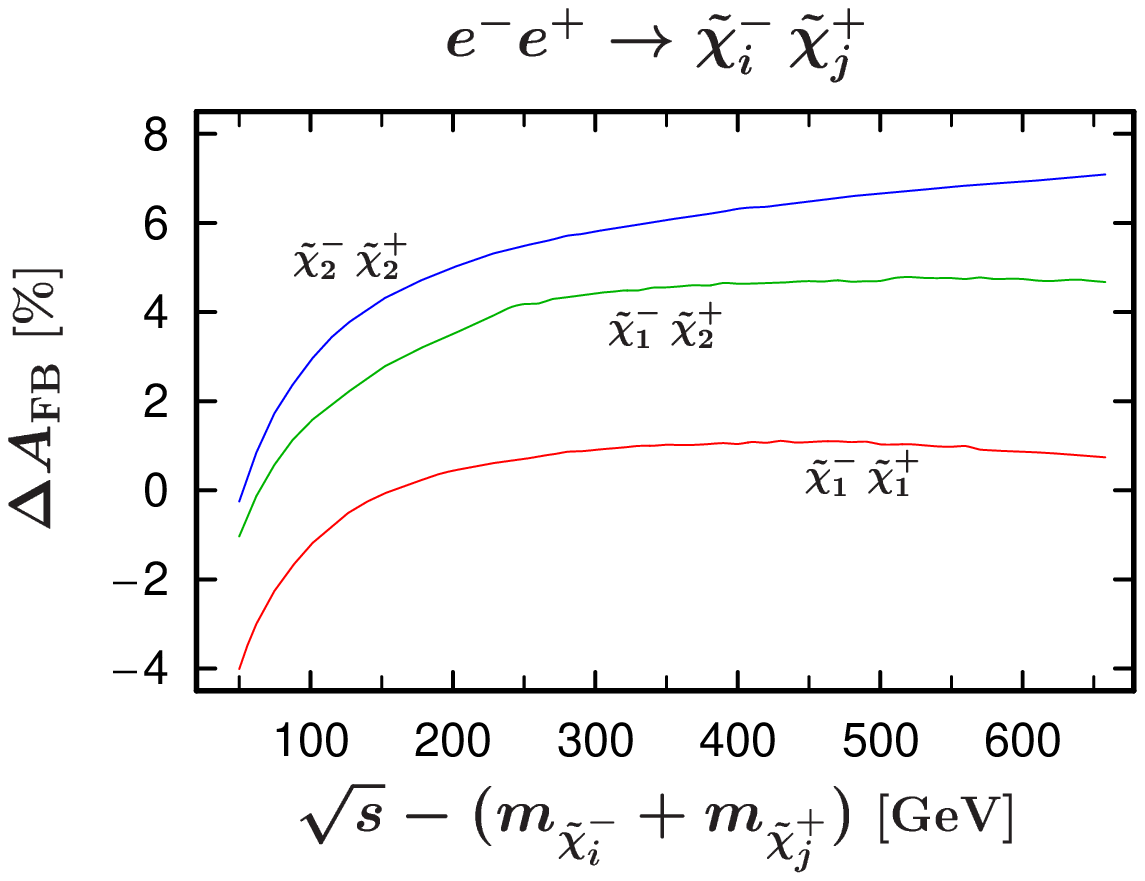}}}}
\end{center}
\caption{Left: Forward-backward asymmetry for chargino pair production at tree-level \{dotted\}, with weak \{dashed\}, and with complete corrections 
\{solid\}. Right: Complete corrections to $A_{\rm FB}$.}
\label{fig:cha_AFB1}
\end{figure}
\begin{figure}[h!]
\begin{center}
\mbox{\mbox{\resizebox{84mm}{!}{\includegraphics{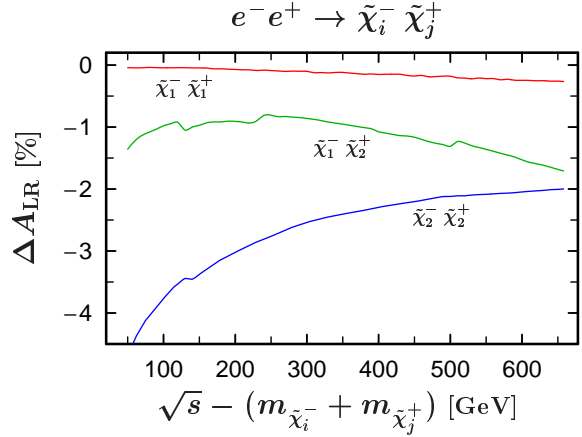}}}\hspace{-5mm}}
\end{center}
\caption{Complete corrections to $A_{\rm LR}$ for chargino pair production.}
\label{fig:cha_ALR1}
\end{figure}
Moreover, we present results for the forward-backward and left-right asymmetry $A_{\rm FB}$ and $A_{\rm LR}$ of chargino production
in the Figs.~(\ref{fig:cha_AFB1}, \ref{fig:cha_ALR1}).
These are defined by
\begin{eqnarray}
A_{\rm FB} = \frac{\s_{\rm F}-\s_{\rm B}}{\s_{\rm F}+\s_{\rm B}}\,,\hspace{2cm} A_{\rm LR} = \frac{\s_{\rm L}-\s_{\rm R}}{\s_{\rm L}+\s_{\rm R}}\,,
\end{eqnarray}
where $\s_{\rm F}\equiv \s(\cos\t_{\vec{p}_1\vec{k}_1}\geq 0)$ with $\t_{\vec{p}_1\vec{k}_1}$ is the angle between the incoming electron and the outgoing
$\nch_i$ in the CMS frame. $\s_{L/R}$ denotes the total cross section for left/right-handed electrons and unpolarized positrons.
Both asymmetries obtain sizeable corrections for all three production channels. The kinks in the lines can be traced back
to so-called normal and anomalous thresholds.

\section{Conclusions}\label{Conclusions}
We have presented in detail the calculation of $O(\a)$ radiative corrections to the pair production of charginos and neutralinos within the
Minimal Supersymmetric Standard Model.
We discussed a possible separation of weak and QED corrections. Although only the sum of virtual and real bremsstrahlung corrections 
have physical meaning, a separation of these two contributions has its advantages.
The QED corrections are treated numerically by integration over the phase space with an additional photon in the final state.
Quite often a cancellation between QED and weak corrections occurs leading to a small total correction.
Furthermore, we used the structure function formalism to include higher order QED corrections and to split the QED corrections into
a universal and a non-universal part. The non-universal QED corrections contain the integration over the three particle final state
and are in the case of the total cross-section often negligible.
The numerical analysis is based on the SPS1a' benchmark scenario, proposed within the SPA project.
The $\drbar$ parameters of the SPA convention are translated into a set of on-shell parameters, which serve as input for our on-shell
renormalization scheme. Such translation tools make it possible to compare cross sections, masses, etc. calculated in different
renormalization schemes.
The presented numerics shows results for the total cross-section, left-right, and forward-backward asymmetries.
In all three cases we have sizeable radiative corrections around the 10\% and in some cases even higher.
\section*{Acknowledgements}
We thank for useful correspondence with W.~Porod and T.~Fritzsche.
The authors acknowledge support from EU under the
HPRN-CT-2000-00149 network programme. The work was also supported
by the ``Fonds zur F\"orderung der wissenschaftlichen Forschung'' of Austria,
project no. P13139-PHY.

\end{document}